\documentclass[12pt]{JHEP}

\usepackage{amsfonts}
\usepackage{amssymb}
\usepackage{amsmath2000}
\usepackage{epsfig}

\newcommand{\be}{\begin{equation}}
\newcommand{\ee}{\end{equation}}
\newcommand{\bq}{\begin{eqnarray}}
\newcommand{\eq}{\end{eqnarray}}

\title{ A new analytic approach to physical observables in QCD }

\author{  Wajdi Gaddah \\Centre for Particle Theory
\\Department of Mathematical Sciences\\University of
Durham\\South Road\\Durham, DH1 3LE, England\\E-mail:
\email{W.A.Gaddah@durham.ac.uk}  }

\abstract{An analytic ghost-free model for the QCD running coupling
$\alpha(Q^2)$ is proposed. It is constructed from a more general  
approach we developed particularly for investigating physical observables 
of the type $F(Q^2)$ in regions that are inaccessible to perturbative 
methods of quantum field theory. This approach directly links the 
infrared (IR) and the ultraviolet(UV) regions together under the causal 
analyticity requirement in the complex $Q^2-$plane. Due to the inclusion 
of crucial non-perturbative effects, the running coupling in our model
not only excludes unphysical singularities but also freezes to a finite
value at the IR limit $Q^2=0$. This makes it consistent with a popular 
phenomenological hypothesis, namely the IR freezing phenomenon.   
Applying this model to compute the Gluon condensate, we obtain
a result that is in good agreement with the most recent
phenomenological estimate. Having calculated the $\beta-$function
corresponding to our QCD coupling constant, we find that it behaves
qualitatively like its perturbative counterpart, when calculated beyond
the leading order and with a number of quark flavours allowing for the 
occurrence of IR fixed points. }

\keywords{QCD, Running coupling constant, $\beta-$function and 
Gluon condensate}

\preprint{DCPT-02/39 }

\begin{document}

\section{Introduction}

A fundamental problem in the theory of particle physics lies in the 
description of hadron interactions in the infrared region. The property
of asymptotic freedom \cite{Gross73} in QCD allows us to investigate 
the interactions of quarks and gluons at short distances using the
standard perturbation theory. However, there is a number of 
phenomena whose description is intractable in perturbation theory, for 
example quark confinement and gluon and quark condensates. Hence, the 
search for calculational techniques that go beyond conventional 
perturbation theory remains essential (not only in QCD but in other 
field theories as well).

In standard perturbative QCD, for any number of quark flavours $n_f$,
the running coupling constant (to leading order):
\be 
\alpha^{(1)}_{\mathrm{PT}}(Q^2)=\frac{4\pi}{\beta_0}\,
\frac{1}{\ln(Q^2/\Lambda^2)} \qquad \textrm{with}\qquad
\beta_0=11-\frac{2}{3}n_f\,,
\label{1.1}
\ee 
diverges at a small mass scale $Q=\Lambda$, creating the so-called Landau 
ghost-pole problem. Taking next loop corrections into account does not 
alter the essence, and leads only to additional branch cuts along the 
positive real axis $\mathrm{Re}\{Q^2 \}>0$. This problem prevents the 
use of perturbative 
expansion at small momentum transfers $Q\thicksim\Lambda$ and, in addition, 
generates infrared (IR) renormalon singularities on the positive real axis
of the Borel parameter, destroying attempts to sum up the perturbative 
series \cite{Parisi78,Simonov93,Badalian97,Grunberg96}. Generally speaking, 
as $Q^2$ comes below or near $\Lambda^2$ , non-perturbative effects become 
the most dominant and the perturbative expansion becomes useless.

Another indication that the perturbative formalism is incomplete, 
and cannot describe the low energy physics unless it is supplemented by 
non-perturbative corrections, comes from considering its analyticity 
structure in the complex $Q^2$-plane \cite{Gardi98}; the upshot being  
that any QCD observable, which depends on a spacelike \footnote{In this 
paper, the metric with signature (-1,1,1,1) is used so that $Q^2 >0$ 
corresponds to a spacelike (Euclidean) 4-momentum transfer.}  
momentum variable $Q^2$, is expected to be an analytic function of $Q^2$ 
in the entire complex plane except the negative real (timelike) axis. 
Singularities on the timelike axis are meaningful since they correspond to 
production of on-shell particles, while their existence on the spacelike 
axis is  non-physical as this would violate causality (i.e. causal 
analyticity structure) \cite{Gardi98,Webber98}.
For example, if a generic QCD observable were calculated to leading order, 
it would depend on $\alpha^{(1)}_{\mathrm{PT}}(Q^2)$, inheriting the 
Landau-singularity at $Q^2=\Lambda^2$. This singularity is obviously 
non-physical as it lies on the positive real axis $\mathrm{Re}\{Q^2\}>0$.
From this point of view, causality constrains the $Q^2-$dependence of
physical observables, and is consistent with perturbative results only if 
the coupling constant is non-singular on the entire $Q^2$-plane with the 
exception of the timelike axis $\mathrm{Re}\{Q^2\}<0$.

A small number of models for the QCD running coupling $\alpha(Q^2)$ have 
been proposed \cite{Grunberg96,Webber98,Shirkov97,Troyan96,Alekseev97} to 
comply with the causality condition, i.e. the requirement
of analyticity in $Q^2$. However, the issue of which of these is the most 
realistic is still a moot point. Quite recently, a good attempt has been 
made, by Shirkov and Solovtsov \cite{Shirkov97}, to devise a model for the 
running coupling that is completely free of singularities in the IR region. 
In this approach, the analytic running coupling is defined via the 
so-called K\"all\'en-Lehmann spectral representation as: 
\be 
\alpha_{\mathrm{an}}^{(n)}(Q^2)=4\,\int^{\infty}_{0} 
\frac{\rho^{(n)}(x)}{(x+Q^2-i\epsilon)}\, dx\,,
\label{1.2}
\ee
with the $n$-loop spectral density $\rho^{(n)}(x)$ given by
\be 
\rho^{(n)}(x) = \frac{1}{4 \pi}\mathrm{Im}\{\alpha^{(n)}_{\mathrm{PT}}
(-x-i\epsilon)\}\,,
\label{1.3}
\ee
where $\alpha^{(n)}_{\mathrm{PT}}$ is the coupling constant obtained from 
perturbation theory (PT) at the n-loop approximation. To leading order, 
the corresponding spectral density reads as:
\be 
\rho^{(1)}(x)=\frac{\pi}{\beta_0} \,
\frac{1}{[\pi^2+\ln^2(x /\Lambda^{2})]}\,.
\label{1.4}
\ee
Inserting this into (\ref{1.2}) gives the one-loop analytic spacelike 
coupling constant:
\be 
\alpha^{(1)}_{\mathrm{an}}(Q^2)=\frac{4\pi}{\beta_0}\,
\bigg[\frac{1}{\ln(Q^2/\Lambda^2)}
+\frac{1}{1-Q^2/\Lambda^2}\bigg]\,.
\label{1.5}
\ee 
This expression is consistent with the causality condition.  
The first term on the right-hand side of (\ref{1.5}) preserves the standard 
UV behaviour whereas the second term compensates for the ghost-pole at 
$Q^2=\Lambda^2$. Obviously the second term in (\ref{1.5}) comes from the 
spectral representation and enforces the required analytic properties. 
Thus, it is essentially non-perturbative. However, as mentioned in 
Ref.\cite{Webber98}, as this term introduces a $1/Q^2$ correction to
$\alpha^{(1)}_{\mathrm{PT}}(Q^2)$ at large $Q^2$ it would make (\ref{1.5}) 
unsuitable as an input quantity for observables that are proportional 
to the running coupling (in leading order) but are not expected to 
have $1/Q^2$ corrections (such as the $e^+ e^-$ total cross section) 
because in such cases the unwanted $1/Q^2$ correction would have to be
artificially cancelled.
By construction, (\ref{1.5}) does not include any adjustable parameters 
other than the QCD characteristic mass scale $\Lambda$. Thus, one might 
prefer to have a model with free extra parameters that would allow for the 
fitting of the coupling to lower energy experimental data and hence
further improve perturbative results at low $Q^2$. An example of a model of
this type is suggested in Ref.\cite{Troyan96}, where the one-loop coupling
constant assumes the form:
\be 
\alpha^{(1)}(Q^2)=\frac{4\pi}{\beta_0}\,
\bigg[\frac{Q^{2\,p}/\Lambda^{2\,p}}{Q^{2\,p}/\Lambda^{2\,p}+C_p}\bigg] 
\;\frac{p}{\ln(Q^{2\,p}/\Lambda^{2\,p}+C_p)}\,,
\label{1.6}
\ee 
with $C_p\ge 1$. For $Q^2\gg \sqrt[p]{C_p}\,\Lambda^2$, this expression
recovers the perturbative asymptotic form (\ref{1.1}). Although (\ref{1.6})
is a ghost-pole free expression, being analytic for all $Q^2\in [0,\infty)$
, it contains unphysical singularities on the complex $Q^2-$plane (for all 
$p>1$) and hence does not comply with causality. A careful analytic study
of non-perturbative effects is always important in QCD because it provides 
more reliable and useful information about the IR region, which is 
inaccessible to perturbative methods.

The main purpose of this paper is to construct a new analytic approach 
based on a different and more general methodology to that of Shirkov 
and Solovtsov, which improves perturbative results outside the asymptotic 
domain and respects the causal analyticity structure in, at least, the
right half of the complex $Q^2-$plane.
Having achieved this, we apply our method to the QCD coupling constant 
to solve the Landau 
ghost-pole problem without altering the correct perturbative behaviour in 
the UV region. Then, we show that the resultant QCD coupling in this 
way freezes to a finite value at the origin (i.e. $Q^2=0$), supporting 
the IR freezing idea which has long been a popular and successful 
phenomenological hypothesis \cite{Mattingly94,Stevenson94}. Moreover, 
from the viewpoint of the new background field formalism 
\cite{Simonov93,Badalian97} our approach provides better estimates 
for the IR fixed points than that of Shirkov and Solovtso.

The layout of this paper is as follows: 
In section 2, we discuss our analyticization procedure on general grounds,
proposing a new integral representation for physical observables that
depend on a spacelike momentum variable (squared) $Q^2$. In the next
section, we present a full derivation of our analytic coupling constant,
giving a new expression that depends on a free extra parameters
$\lambda_e$ which can be used to tune the coupling to lower energy
experimental data. An approximation scheme for estimating the value 
of $\lambda_e$ is included in section 4.
In section 5, we carry out a comprehensive comparison between the 
predictions in our approach and those estimated by other theoretical 
methods, which include conventional perturbation theory, optimized
perturbation theory, background field formalism and the analytic approach 
of Shirkov and Solovtsov. In this section, we test our model on
a fit-invariant IR characteristic integral extracted from jet physics 
data. For further applications, we use our analytic coupling constant in the
instanton density derived from the dilute instanton-gas approximation
to estimate the gluon condensate. Then, we show that the result obtained
is in good agreement with the value phenomenologically estimated from
the QCD sum rules. Finally, we end this section by calculating 
the $\beta-$function which corresponds to our analytic coupling 
constant and compare its behaviour with the perturbative counterpart. 
We further show that the $\beta-$function in our approach and that 
in higher-loop order perturbation theory behave qualitatively 
the same in the range of $n_f-$values that allows perturbation 
theory to obtain IR fixed points.   
The last section is devoted to our conclusions.

\section{Analyticization Procedure}

We will illustrate our method with the following pedagogical example.
Consider a physical observable $F$ depending on a positive variable $q$ 
with the dimensions of energy squared. Then, it follows from the principle 
of causality, mentioned earlier, that $F(q)$ can be analytically continued 
to the complex plane excluding the negative real axis. 
As we shall show, this will allow us to reconstruct $F(q)$ from its high
energy behaviour by using the contour integral representation: 
\be  
\bar F(q,\lambda)=\frac{1}{2\pi i}\, \int_{C_{UV}}
\, \,\frac{{\mathrm e}^{\lambda(k-q)}}{k-q}\;\,F(k) \:dk\, ,
\label{2.1}
\ee
where $C_{UV}$ is a very large incomplete circle, in the complex $k-$plane,
with radius $R_{UV}$ and centre at the origin, beginning just below the
negative real axis and ending just above and $q \in \{\;\textrm{Re}\;k\,
:\:\:0<\textrm{Re}\;k<R_{UV}\}$ as depicted in Fig.(\ref{Fig1}).
\FIGURE{ \unitlength 1cm 
\resizebox{14cm}{10.2cm}{\epsfig{file=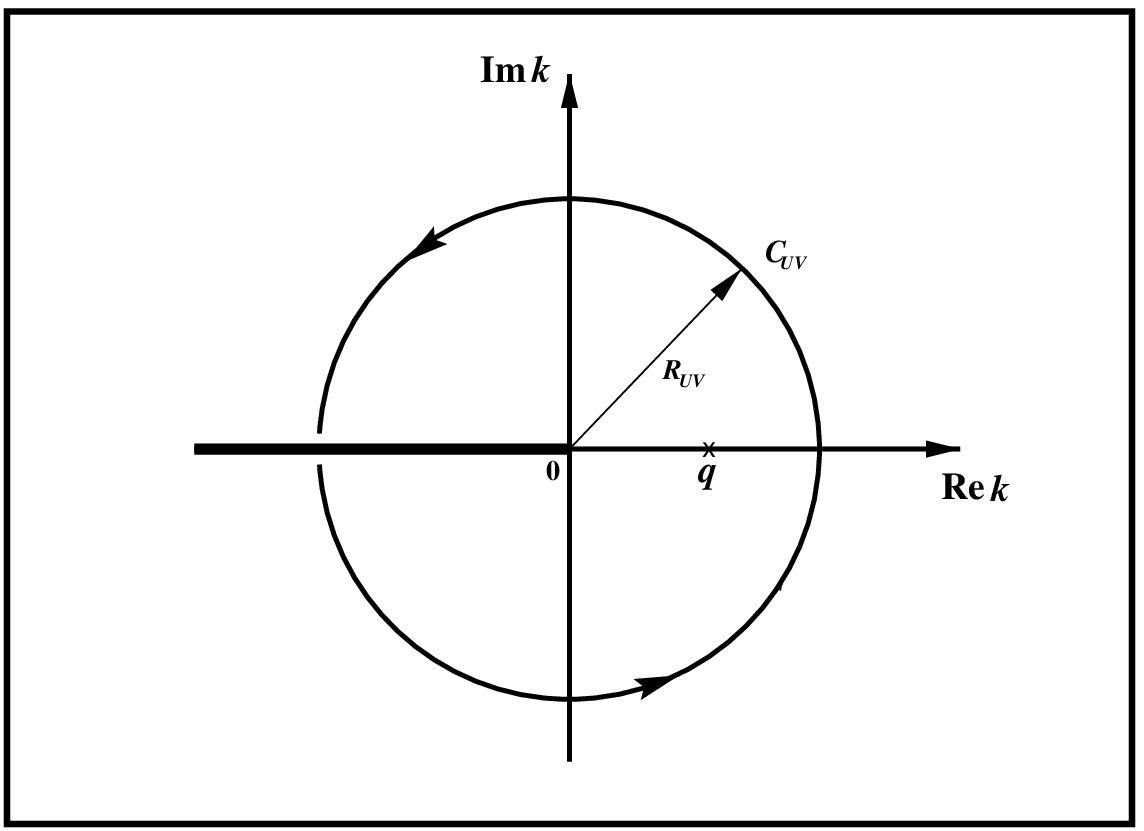}} 
\caption{Sketch of the integration contour $C_{UV}$ in the cut complex 
$k-$plane. The thick line represents a branch cut and the cross indicates 
the location of $q$.} 
\label{Fig1} }

We will show that $F(q)$ is recovered as the limit as $\lambda\rightarrow
\infty$ of $\bar F(q,\lambda)$, and this depends mainly on the ultraviolet 
behaviour of $F(q)$  when we take $R_{UV}$ to be sufficiently
large. In order to achieve this goal, we exploit the analyticity of
$F(k)$ in the complex $k$-plane, using Cauchy's theorem to replace the 
UV-boundary $C_{UV}$ in (\ref{2.1}) with a small circle $C_0$ around $q$ 
and a keyhole shaped contour $C_k$, depicted in Fig.(\ref{Fig2}), 
surrounding the cut and not connected to $C_0$.
\FIGURE[ht]{ \unitlength 1cm
\resizebox{14.7cm}{7.2cm}{\epsfig{file=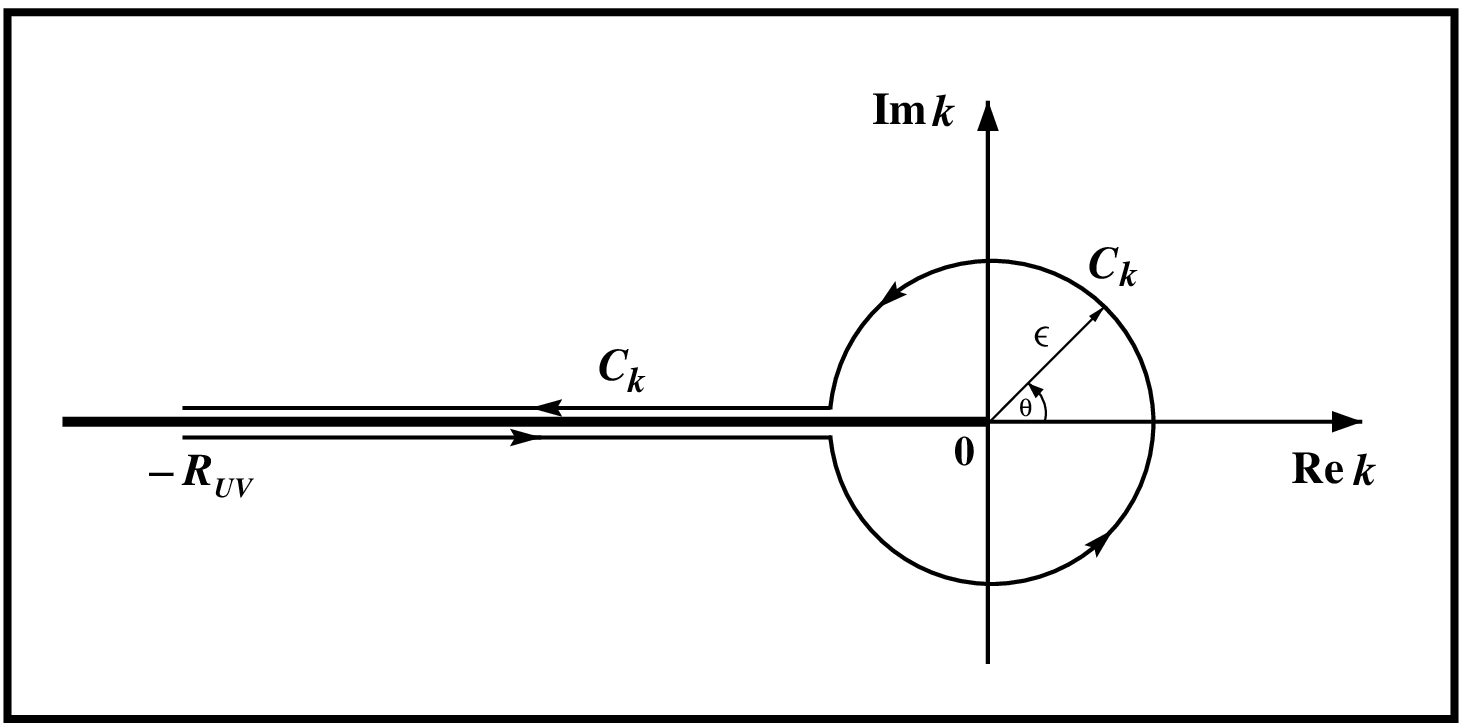}} 
\vspace{-0.13in}
\caption{Sketch of the keyhole shaped contour $C_k$. 
The thick line represents the branch cut $\textrm{Re}\;k\le 0$.
Here, $\epsilon$ is an infinitesimally small radius 
and $\theta\in(-\pi,\pi]$.}
\label{Fig2} }
For this deformed contour, Cauchy's integral formula over 
the circle $C_0$ yields $F(q)$ which together with the 
integral over the keyhole contour $C_k$ transforms 
(\ref{2.1}) into the form:
\be 
\bar F(q,\lambda)=\,F(q) \,+\,I_{C_k},
\label{2.2}
\ee  
where 
\be 
I_{C_k}=\frac{1}{2\pi i}\, \int_{C_k}
\, \,\frac{{\mathrm e}^{\lambda(k-q)}}{k-q}\,\,F(k) \, dk.
\label{2.3}
\ee
If we restrict attention to those functions $F(k)$ which 
are finite on the line of discontinuity along the
negative real $k$-axis, e.g. $F(k)\propto \sqrt{k}$,
then we can write $I_{C_k}$ more explicitly as:
\be 
I_{C_k}=\int^{R_{UV}}_{\epsilon}
\, \,\frac{{\mathrm e}^{-\lambda(r+q)}}{r+q}\;\;\Delta(r)\,dr
\;+\;\frac{\epsilon \, {\mathrm e}^{-\lambda q}}{2 \pi}
\,\int^{\pi}_{-\pi}{\mathrm e}^{\lambda \epsilon \cos(\theta)}
\,\;\textrm{f}_{\epsilon}(\theta,q)\;d\theta,
\label{2.4}
\ee
with
\be 
\Delta(r)=\frac{1}{2\pi i}\;[\; F(r\,{\mathrm e}^{i\pi})\,- \,
F(r\,{\mathrm e}^{-i\pi})\;]\,, 
\label{2.5}
\ee
and
\be 
\textrm{f}_{\epsilon}(\theta,q)=\frac{1}{2}\, \Bigg[\,
\frac{{\mathrm e}^{i\lambda \epsilon \sin(\theta)}\;
F(\epsilon {\mathrm e}^{i\theta})}{\epsilon-q\;{\mathrm e}^{-i\theta}}+ 
\frac{{\mathrm e}^{-i\lambda \epsilon \sin(\theta)}\;
F(\epsilon {\mathrm e}
^{-i\theta})}{\epsilon-q\;{\mathrm e}^{i\theta}} \,\Bigg].  
\label{2.6}
\ee 
Owing to the discontinuity of $F(k)$ across the cut, given by 
$2\pi i\, \Delta(|k|)\,$, the sum of the line integrals on each 
side of the branch cut does not cancel. Since the region of 
analyticity of $F(k)$:
\be
S_{an}=\{k=r\,{\mathrm e}^{i \theta}:\,r>0\,,\,-\pi<\theta \le \pi \},   
\label{2.7}
\ee 
includes a portion of the real axis on which $F(k)$ is real,
it follows that $F(k)$ satisfies the Schwarz Reflection 
Principle \cite{Wyld99}:
\be
F^{*}(k)=F({k}^{*}).
\label{2.8}
\ee 
This allows us to rewrite $\Delta(r)$ and 
$ \textrm{f}_{\epsilon}(\theta)$ in the more compact forms:
\be 
\Delta(r)=\frac{1}{\pi}\;\textrm{Im} 
\Big\{ \; F(r\,{\mathrm e}^{i\pi})\;\Big\}\,, 
\label{2.9}
\ee
and
\be 
\textrm{f}_{\epsilon}(\theta,q)=\textrm{Re}\, \Bigg\{\,
\frac{{\mathrm e}^{i\lambda \epsilon \sin(\theta)}\;
F(\epsilon\, {\mathrm e}^{i\theta})}
{\epsilon-q\;{\mathrm e}^{-i\theta}}\,\Bigg\},  
\label{2.10}
\ee 
which imply that $I_{C_k}$ is a purely real quantity
as it should be.

In the following, we will investigate the behaviour of 
$I_{C_k}$ at sufficiently large values of $\lambda$. 
Since $F(r\,{\mathrm e}^{i\theta})$ is assumed finite 
throughout the whole region $S_{an}$, we take its magnitude
to be bounded on the infinitesimal circle $|k|=\epsilon$, see
Fig.(\ref{Fig2}), such that:
\be
|F(\epsilon\,{\mathrm e}^{i\theta})|\le M\,, 
\qquad \textrm{for}\quad -\pi<\theta\le\pi\,,
\label{2.11}
\ee
where $M$ is a positive constant which may depend on $\epsilon$. 
Using this to estimate the upper bound of 
the second term in (\ref{2.4}), we find that:
\be
\frac{\epsilon \,{\mathrm e}^{-\lambda q}}{2 \pi}\,
\Bigg |\int^{\pi}_{-\pi}{\mathrm e}^{\lambda \epsilon 
\cos(\theta)}\,\;\textrm{f}_{\epsilon}(\theta,q)\;d\theta 
\Bigg | \le \frac{\epsilon\,M\,{\mathrm e}^
{-\lambda(q-\epsilon)}}{q-\epsilon}. 
\label{2.12}
\ee
If we consider only those functions $F$ for which $M$ can be chosen 
such that $\displaystyle{\lim_{\epsilon\rightarrow 0}}\;\epsilon M=0$ 
then the second term in (\ref{2.4}) will vanish in this limit and 
accordingly~$I_{C_k}$ reduces to the form:
\be
I_{C_k}(q,\lambda)=\int^{R_{UV}}_{0}
\,\,\frac{{\mathrm e}^{-\lambda(r+q)}}{r+q}\;\,\Delta(r)\; dr,
\qquad q>0\,, 
\label{2.13}
\ee 
which is more suitable for further calculations. 
For the case under consideration, since $\Delta(r)$ has an upper 
bound $N>0$, i.e. $|\Delta(r)|\le N\,$, in the interval 
$0\le r \le R_{UV}$, we deduce that:
\be
|I_{C_k}|\le N\;\frac{{\mathrm e}^{-\lambda q}}{\lambda q} 
\;\left(1-{\mathrm e}^{-\lambda R_{UV}}\right). 
\label{2.14}
\ee 
So for large values of $\lambda$, the function $I_{C_k}(q,\lambda)$ 
decays rapidly to zero, and: 
\be
\lim_{\lambda \rightarrow \infty} I_{C_k}(q,\lambda)=0\,.
\label{2.15}
\ee 
In general, (\ref{2.15}) remains true even if the finiteness 
restriction on $F(k)$ along the negative real $k$-axis,
together with the constraint
$\displaystyle{\lim_{\epsilon\rightarrow 0}\epsilon M=0}$,
is not fulfilled.
For example, if $F(k)$ has a finite number of singularities on 
the negative real axis then we can avoid these points by keeping
the two edges of the keyhole contour $C_k$ a small distance $\delta$
away from the cut which they surround. In this way, we can rewrite
(\ref{2.3}) as:
\be 
I_{C_k}=\int^{R_{UV}}_{\epsilon'}
{\mathrm e}^{-\lambda(x+q)}\;\;
\textrm{g}(x,q,\lambda)\,dx
\,+\,\frac{\epsilon\,{\mathrm e}^{-\lambda q}}{2\pi}\,
\int^{\pi-\eta}_{-\pi+\eta}
{\mathrm e}^{\lambda \epsilon \cos(\theta)}\,\;
\textrm{f}_{\epsilon}(\theta,q)\;d\theta,
\label{2.15b}
\ee
where $\epsilon'=\sqrt{\epsilon^2-\delta^2}\;$ with  
$\epsilon >\delta\,$, $\,\eta=\arctan{(\delta/\epsilon')}$ 
and:
\be
\textrm{g}(x,q,\lambda)=\frac{1}{\pi}\;\mathrm{Im}\,\Big{\{}\,
\frac{\mathrm{e}^{i\lambda\,\delta}}{x+q-i\delta}\;\;
F(-x+i\delta)\,\Big{\}}\,.
\label{2.15c}
\ee
Since $F$ is bounded on the contour $C_k\,$, we have:
\be
|I_{C_k}|\le \,\frac{N'}{\pi}\;\frac{{\mathrm e}^{-\lambda q}}
{\lambda(q+\epsilon'-\delta)}\; 
\Big[{\mathrm e}^{-\lambda\epsilon'}
-{\mathrm e}^{-\lambda R_{UV}}\Big]\;+\;
\frac{\epsilon\,M\,(\pi-\eta)}{\pi}\;\;
\frac{{\mathrm e}^{-\lambda(q-\epsilon)}}{q-\epsilon}\,, 
\label{2.15d}
\ee 
where $N'$ is the upper bound of $F(-x+i\delta)$ for all
$x\in[\epsilon',R_{UV}]\,$. As both $N'$ and $M$ are 
$\lambda$-independent, in the limit as $\lambda\to\infty$
we obtain (\ref{2.15}) as claimed above.
So, by taking the limit as $\lambda\to\infty$ in (\ref{2.2})
we deduce from (\ref{2.15}) that:
\be
\lim_{\lambda \rightarrow \infty} \bar F(q,\lambda)=F(q)\,.
\label{2.16}
\ee

To have some notion of the functional dependence of $\bar{F}(q,\lambda)$
on $\lambda$, for a fixed $q>0$, consider the simple example where 
the imaginary part of $F(k)$ assumes a constant value $\mu_0$ along 
the upper edge of the cut, e.g. $F(k)\propto \ln k$.
In this case, we have $\Delta(r)=\mu_0/\pi$ and
a straightforward integration of (\ref{2.13}) yields: 
\be
I_{C_k}(q,\lambda)=\,\frac{\mu_0}{\pi}\;
\Big[E_1(\lambda q)-E_1(\lambda(q+R_{UV}))\Big] \,,
\label{2.17}
\ee 
where $E_1(x)$ denotes the exponential integral function \cite{Arfken85}
defined by:
\be
E_1(x)=\int^{\infty}_{x}\frac{{\mathrm e}^{-t}}{t}\:dt=-\gamma-\ln(x)-
\sum_{n=1}^{\infty}\frac{(-1)^n}{n!\,n}\;x^n \qquad \textrm{for}\:\:\:x>0\,,
\label{2.18}
\ee
here $\gamma \cong 0.577215$ is the well-known Euler's constant.
By inserting (\ref{2.17}) into (\ref{2.2}) and sending
$R_{UV} \to \infty$, we immediately arrive at an explicit solution:
\be
\bar F(q,\lambda)=F(q)+\,\frac{\mu_0}{\pi}\;E_1(\lambda q)\,.
\label{2.19}
\ee 
As illustrated by the graph in Fig.(\ref{Fig3}), the variation of $\bar 
F(q,\lambda)$ with respect to $\lambda$ depends largely on both the sign 
of $\mu_0$ and the behaviour of $ E_1(\lambda q) $. With increasing 
$\lambda$, we see that $\bar F(q,\lambda)$ evolves from $-\infty$, 
for $\mu_0<0$, or $+\infty$, for $\mu_0>0$, in a continuous
progression until it settles down to the value $F(q)$ at large 
values of $\lambda$. Moreover, we discover  from (\ref{2.19}) 
that $\bar F(q,\lambda)$ converges to $F(q)$ faster as $q$ increases.
\FIGURE{ \unitlength 1cm 
\resizebox{13cm}{14cm}{\epsfig{file=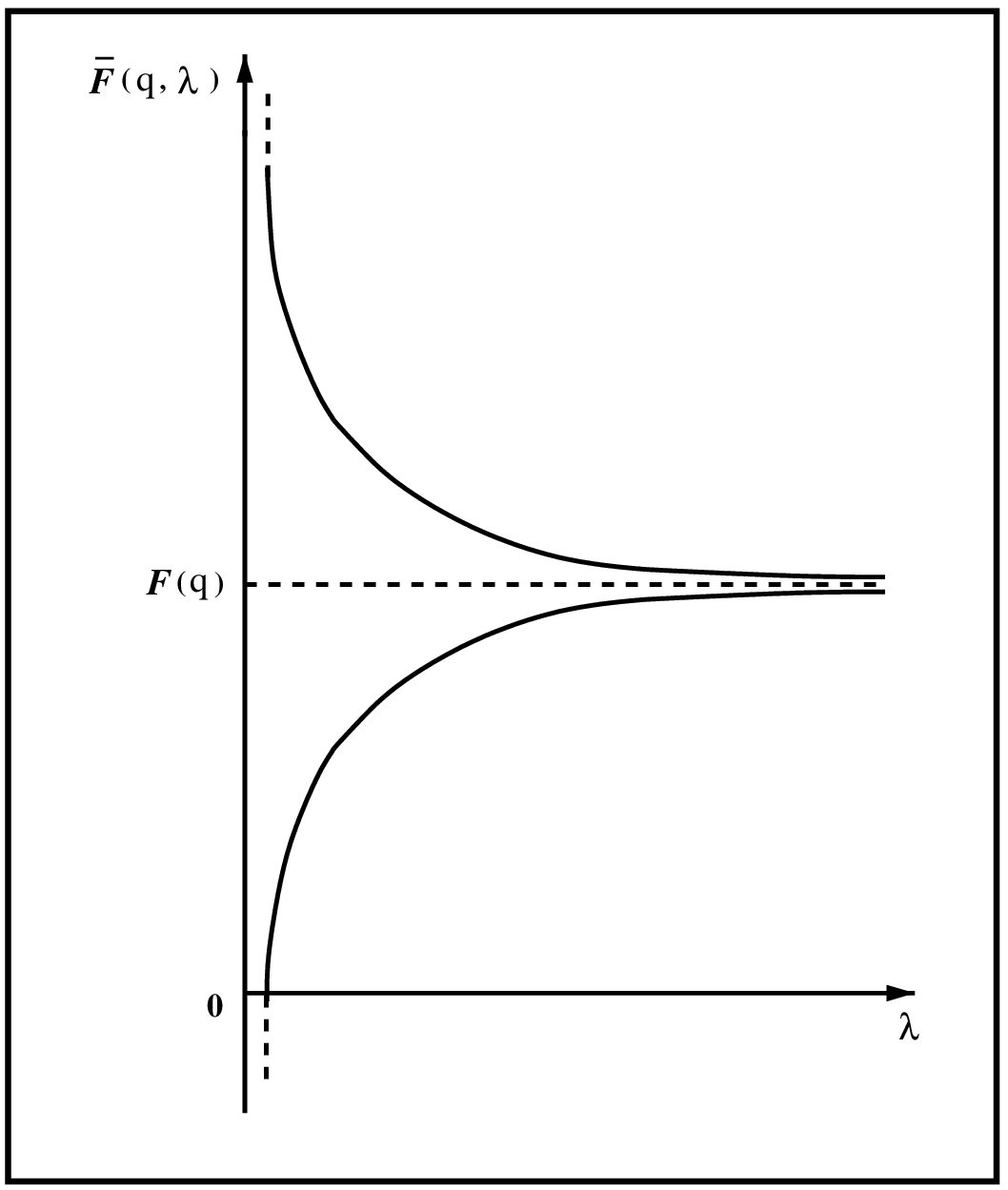}} 
\caption{The variation of $\bar F(q,\lambda)$ versus 
$\lambda$ at a fixed value $q$\,. 
The lower and the upper curves describe the cases 
in which $\mu_0<0$ and $\mu_0>0$, respectively. 
The dashed line represents $F(q)$.}
\label{Fig3} }

To summarise, by exploiting the analyticity of $F(k)$ in (\ref{2.1}),
we can obtain $F(q)$ as $\lim_{\lambda\rightarrow \infty} 
\bar F(q,\lambda)$, and if we take the radius $R_{UV}$ of 
the contour $C_{UV}$ to be sufficiently large the integrand depends 
on the large $q$ behaviour of $F$. So far, our consideration 
has been restricted to the case in which $F(q)$, as a true function for the 
observable in question, complies with the principle of causality. In practice, 
however, we do not know the function $F(q)$ exactly, even for large $q$, but 
for a theory with asymptotic freedom we can approximate the large $q$
behaviour of $F(q)$ using perturbation theory for $q$ in some high energy 
domain $D_{PT}$ so that $F_{PT}(q)\cong F(q)$ for $q\in D_{PT}$. Thus, for
large enough $R_{UV}$, we can safely use $F_{PT}(k)$ as a reasonable 
substitute for $F(k)$ in our contour integral (\ref{2.1}), where now $k$ is 
any point on $C_{UV}$. This gives the approximation $\bar F_{PT}(q,\lambda)
\approx \bar F(q,\lambda)$, but taking $\lambda$ to infinity in this would
simply reproduce the perturbative estimate for $F(q)$, leading to nothing new. 
Now $F_{PT}(q)$, as a perturbatively calculated quantity, often suffers from 
a serious defect by having singularities in the complex plane away from the 
negative real axis. 
Good examples of this are the effective coupling constant and the running 
mass in perturbative QCD. These extra singularities are unphysical and 
do not exist in the true function $F(q)$. Hence, they should be eliminated 
in our approximation in order to reinstate causality. So, if instead 
of using the limiting value of $\bar F_{PT}(q,\lambda)$ we were to use a
finite value of $\lambda$, say $\lambda_0$, then we might have an expression 
in which these unphysical singularities are absent and we will give
an example in which this happens.
Furthermore, taking a finite value of $\lambda$ in $\bar{F}_{PT}(q,\lambda)$ 
can give an improvement over the perturbative estimate in favourable 
circumstances for the following reason: 
Viewed as a function of $\lambda$ with $q$ held fixed, 
$\bar{F}_{PT}(q,\lambda)$ is a better estimate of the true function $\bar
F(q,\lambda)$ for small $\lambda$ than it is for large $\lambda$ since if 
we scale the integration variable in (\ref{2.1}) then:
\be
\bar F_{PT}(q,\lambda)=\frac{1}{2\pi i}\,  \int_{|k|=\lambda R_{UV}}
\, \,\frac{{\mathrm e}^{(k-\lambda q)}}{(k-\lambda q)}\,\,F_{PT}(k/\lambda) 
\, dk\, .
\label{2.20}
\ee
For large enough $R_{UV}$ we can replace the contour of integration in
(\ref{2.20}) by $|k|=R_{UV}$ since the difference this makes is a small
contribution from the negative real axis:
\be
\delta_0=\frac{1}{\pi}\;\Bigg | \int^{R_{UV}}_{\lambda\,R_{UV}}
\, \,\frac{{\mathrm e}^{-(k+\lambda\,q)}}{(k+\lambda\,q)}\;
\textrm{Im} \Big\{\,F_{PT}(k \,{\mathrm e}^{i\pi}/\lambda) 
\, \Big\}   
\; dk\, \Bigg |\,,
\label{m}
\ee 
which is exponentially damped. 
Now we can see that the smaller the value of $\lambda$ is the larger is 
the argument of $F_{PT}$ in the integrand of (\ref{2.20}), and for an 
asymptotically free theory this gives a better approximation to $\bar 
F(q,\lambda)$. So, $\bar F_{PT}(q,\lambda)$ interpolates between the 
true quantity $\bar F(q,\lambda)$ at small $\lambda$ and the perturbative 
estimate $\bar F_{PT}(q)$ at large $\lambda$. Under favourable circumstances 
$\lambda$ can be chosen sufficiently small that $\bar F_{PT}(q,\lambda)$ 
provides a good enough approximation to the exact function 
$\bar F(q,\lambda)$, and sufficiently large that we are estimating 
the large $\lambda$ behaviour of the latter, i.e. $F(q)$.

We will now discuss how to choose an appropriate value $\lambda=\lambda_0$ 
for $\bar F_{PT}(q,\lambda)$, depending on the behaviour of $F_{PT}$, and
pinpoint the most favourable situations in which $\bar F_{PT}(q,\lambda_0)$ 
gives a better estimate for $F(q)$ than $F_{PT}(q)$ does. An illustrative 
example is perhaps the best way to convey these ideas. Consider the simple 
case in which $F_{PT}(q)$ suffers only from one infrared singularity at 
$q=\Lambda^{2}$ of the simple pole type:
\be
F_{PT}(q)=\frac{U(q)}{ q-\Lambda^{2}}\, , 
\label{2.21}
\ee
where $U(q)$ is an analytic function of $q$ in the entire complex $q-$plane 
excluding the negative real axis (e.g. the principle value of $\,\log$). 
Then, define the UV region where the perturbation theory of this toy model 
is trustworthy by:
\be
D_{PT}=\{q: q \gg\Lambda^{2}\}\,.
\label{2.22}
\ee
Replacing $F$ and $\bar F$ with $F_{PT}$ and $\bar F_{PT}$, respectively, 
in our integral formula (\ref{2.1}) and then integrating in the usual way, 
we obtain:
\be
\bar F_{PT}(q,\lambda)=F_{PT}(q)
-\frac{U(\Lambda^2)}{q-\Lambda^{2}}\;{\mathrm e}^{-\lambda(q-\Lambda^{2})}
+H(q,\lambda)\qquad \textrm{for}\;\;q \ne \Lambda^2 \, ,    
\label{2.23}
\ee
and
\be
\bar F_{PT}(\Lambda^2,\lambda)=\lambda\,U(\Lambda^2)+U'(\Lambda^2)
+H(\Lambda^2,\lambda)\qquad  \textrm{for}\;\;q = \Lambda^2 \, ,    
\label{2.24}
\ee
where 
\be
H(q,\lambda)=\int^{R_{UV}}_{0}\;
\frac{{\mathrm e}^{-\lambda(r+q)}}{r+q}\;\,\Delta_{PT}(r)\;dr\,,
\qquad q>0\,,
\label{2.25}
\ee
with
\be 
\Delta_{PT}(r)=-\frac{1}{\pi}\;\,
\frac{\textrm{Im} \{ \; U(r\,{\mathrm e}^{i\pi})\; \} }{r+\Lambda^2} \,.
\label{2.26}
\ee  
Here, we have $R_{UV}\gg \Lambda^2$. 
In this model, (\ref{2.23}) allows us to modify the perturbative 
expression $F_{PT}(q)$ through the variable $\lambda$.
Taking $\lambda$ to infinity in (\ref{2.23}) would only 
reproduce $F_{PT}(q)$:
\be 
\lim_{\lambda \rightarrow \infty }\,\bar F_{PT}(q,\lambda)=F_{PT}(q)\,,
\label{2.28}
\ee  
on the other hand, keeping $\lambda$ fixed at a finite value
$\lambda_0$ would change the behaviour of $F_{PT}(q)$, especially 
in the vicinity of $\Lambda^2$, by an infrared correction term:
\be
\Upsilon_{IR}(q,\lambda_0)=-\frac{U(\Lambda^2)}{q-\Lambda^2}\;\;
{\mathrm e}^{-\lambda_0(q-\Lambda^2)}+H(q,\lambda_0)\,,
\label{2.29}
\ee   
which would, in turn, remove the IR singularity in $F_{PT}(q)$.
So, to reinstate the causal analyticity structure violated by the presence 
of the simple pole singularity on the positive real axis, we would simply 
retain $\Upsilon_{IR}(q,\lambda_0)$ in our formulation. For a finite value of
$\lambda$, the continuity of $\bar F_{PT}(q,\lambda)$ at $q=\Lambda^2\,$:  
\be 
\lim_{q \rightarrow \Lambda^2 }\,\bar F_{PT}(q,\lambda)=
\bar F_{PT}(\Lambda^2,\lambda)\, ,  
\label{2.27}
\ee 
follows directly from (\ref{2.23}) and (\ref{2.24}).

Now, we shall study this example further by investigating the possibility of
finding a proper value $\lambda=\lambda_0$ that would allow $\bar F_{PT}(q,
\lambda_0)$ to match the exact value of the observable $F(q)$ at the point 
$q$ in question. At the same time, we shall consider the case in which this
is not possible and show how to make the best choice of $\lambda_0$ that
can improve perturbative predictions. To explore straightforwardly  the 
criterion of selecting $\lambda_0$, at a fixed value of $q$, we shall
simplify our example further by assuming that $\textrm{Im}\,\{U(r {\mathrm e}
^{i\pi})\}= \omega $, with $\omega$ being some real constant. In this case, 
a straightforward calculation of (\ref{2.25}) yields:
\begin{align}
H(q,\lambda)=-\,\frac{\omega}{\pi (q-\Lambda^2)}&
\; \Big\{ {\mathrm e}^{-\lambda(q-\Lambda^2)} \; 
\Big [ \; E_1(\lambda \Lambda^2)
-E_1(\lambda(\Lambda^2+R_{UV}))\;\Big ]
\nonumber\\
&-E_1(\lambda q)+E_1(\lambda(q+R_{UV}))\;\Big\}
\qquad \textrm{for}\;\;q\ne\Lambda^2\,, 
\label{2.30}
\end{align}  
and
\begin{align}
H(\Lambda^2,\lambda)=-\frac{\omega}{\pi}\;
{\mathrm e}^{-\lambda \Lambda^2 }\,\Big \{\;\frac{1}{\Lambda^2}
-&\frac{{\mathrm e}^{-\lambda R_{UV}}}{\Lambda^2+R_{UV}}
-\lambda\, {\mathrm e}^{\lambda \Lambda^2}\; 
\Big [ \, E_1(\lambda\Lambda^2 )
\nonumber\\
&-E_1(\lambda(\Lambda^2+R_{UV}))\Big ] \,\Big \}
\qquad \textrm{for}\;\;q=\Lambda^2\,.
\label{2.31}
\end{align}
Note that in the limit as $q\to\Lambda^2$ equation (\ref{2.30}) 
tends to (\ref{2.31}).
Substituting (\ref{2.30}) into (\ref{2.23}) and then letting 
$R_{UV}\to\infty$, we obtain:
\be
\bar F_{PT}(q,\lambda)=F_{PT}(q) 
-\Big[ U(\Lambda^2)+\frac{\omega}{\pi}\,E_1(\lambda \Lambda^2)
-\frac{\omega}{\pi}\,E_1(\lambda q)\;
{\mathrm e}^{\lambda(q-\Lambda^2)}\Big]
\,\frac{{\mathrm e}^{-\lambda(q-\Lambda^2)}}{q-\Lambda^2}\,,
\label{2.32}
\ee
which together with the first and the second derivatives:
\be
\frac{d}{d\lambda}\bar F_{PT}(q,\lambda)=\Big[U(\Lambda^2)
+\frac{\omega}{\pi}\,E_1(\lambda \Lambda^2)\,\Big]\,{\mathrm e}^
{-\lambda(q-\Lambda^2)}\,,
\label{2.33}
\ee  
and
\be
\frac{d^2}{d\lambda^2}\bar F_{PT}(q,\lambda)=-\frac{\omega}{\pi} \Big\{
(q-\Lambda^2)\Big[E_1(\lambda \Lambda^2)+\frac{\pi}{\omega}
U(\Lambda^2)\Big] 
+\frac{{\mathrm e}^{-\lambda \Lambda^2}}{\lambda}\Big\}\,{\mathrm e}^
{-\lambda(q-\Lambda^2)},
\label{2.34}
\ee  
allows us to deduce and plot the graphs of the possible cases governing 
the variation of $\bar F_{PT}(q,\lambda)$ against $\lambda$. 
These can be classified in the region $q>\Lambda^2$ according to the
signs of the constants $\omega$ and $U(\Lambda^2)$ in the following way:
\begin{enumerate}
\item 
if $\omega>0$ and $U(\Lambda^2)>0$ then:
$$\frac{d}{d\lambda}\bar F_{PT}(q,\lambda)>0 \qquad \textrm{and}\qquad 
\frac{d^2}{d\lambda^2}\bar F_{PT}(q,\lambda)<0 \,,
\qquad \textrm{for}\;\;\lambda>0\,, $$
which imply that with increasing $\lambda$ the curve of 
$\bar F_{PT}(q,\lambda)$ increases towards $F_{PT}(q)$ with
a downward concavity as depicted in Fig.(\ref{Fig4}.1.a) and 
Fig.(\ref{Fig4}.1.b);
\item
if $\omega>0$ and $U(\Lambda^2)<0$ then there is
a local maximum at $\lambda=\lambda_{max}\,$, which can be
deduced from:
\begin{align}
\frac{d}{d\lambda}\bar F_{PT}(q,\lambda)\Big|_{\lambda=\lambda_{max}}
\!\!&=0\;\Longrightarrow\; 
E_1(\lambda_{max}\Lambda^2)=\frac{\pi}{\omega}
|U(\Lambda^2)|\qquad \textrm{and}
\nonumber \\ 
\frac{d^2}{d\lambda^2}
\bar F_{PT}(q,\lambda)\Big|_{\lambda=\lambda_{max}}&<0 \,;
\nonumber
\end{align}
Here, the way in which $\bar F_{PT}(q,\lambda)$ approaches $F_{PT}(q)$ 
is different because $\bar F_{PT}(q,\lambda)$ rises first to the maximum
$\bar F_{PT}(q,\lambda_{max})$ to fall next towards $F_{PT}(q)$
as $\lambda$ increases beyond $\lambda_{max}$ as shown in Fig.(\ref{Fig4}.2.a) 
and Fig.(\ref{Fig4}.2.b);
\item
if $\omega<0$ and $U(\Lambda^2)<0$ then:
$$\frac{d}{d\lambda}\bar F_{PT}(q,\lambda)<0 \qquad \textrm{and}\qquad 
\frac{d^2}{d\lambda^2}\bar F_{PT}(q,\lambda)>0 \,,
\qquad \textrm{for}\;\;\lambda>0\,, $$
which indicates clearly that with increasing $\lambda$ the curve of 
$\bar F_{PT}(q,\lambda)$ decreases towards $F_{PT}(q)$ with
an upward concavity as depicted in Fig.(\ref{Fig5}.3.a) and 
Fig.(\ref{Fig5}.3.b);
\item
if $\omega<0$ and $U(\Lambda^2)>0$ then there is 
a local minimum at $\lambda=\lambda_{min}$, which can be deduced from:
\begin{align}
\frac{d}{d\lambda}\bar F_{PT}(q,\lambda)\Big|_{\lambda=\lambda_{min}}
&=0 \; \Longrightarrow\; E_1(\lambda_{min}\Lambda^2)=\frac{\pi}{|\omega|}
U(\Lambda^2)\qquad \textrm{and}
\nonumber \\ 
\frac{d^2}{d\lambda^2}
\bar F_{PT}(q,\lambda)\Big|_{\lambda=\lambda_{min}}&>0 \,;
\nonumber
\end{align}
Here, $\bar F_{PT}(q,\lambda)$ decreases in the direction of the minimum
$\bar F_{PT}(q,\lambda_{min})$ as $\lambda$ approaches $\lambda_{min}$ 
from the left and begins to rise towards $F_{PT}(q)$ immediately
after $\lambda$ passes $\lambda_{min}$ as shown in
Fig.(\ref{Fig5}.4.a) and Fig.(\ref{Fig5}.4.b).
\end{enumerate}
\FIGURE[ht]{ \unitlength 1cm 
\resizebox{14.5cm}{14cm}{\epsfig{file=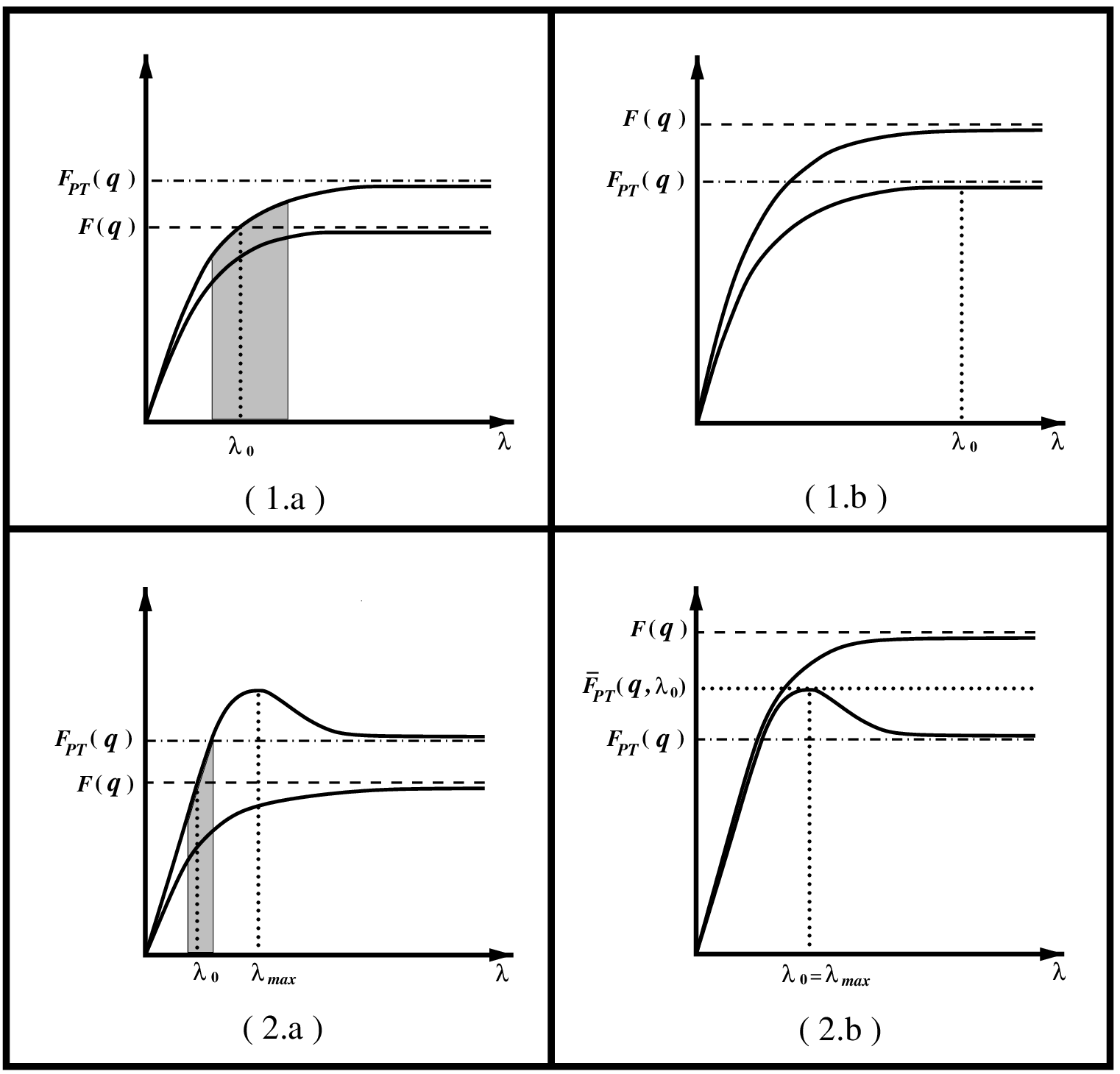}} 
\caption{The variation of $\bar F_{PT}(q,\lambda)$ and the expected 
$\bar F(q,\lambda)$ versus $\lambda$ at a fixed value $q>\Lambda^2$.
In (1.a) and (2.a) the upper and the lower curves describe 
$\bar F_{PT}(q,\lambda)$ and $\bar F(q,\lambda)$ respectively. 
In (1.b) and (2.b) the upper and the lower curves describe 
$\bar F(q,\lambda)$ and $\bar F_{PT}(q,\lambda)$ respectively. }
\label{Fig4} }
\FIGURE[ht]{ \unitlength 1cm 
\resizebox{14.5cm}{14cm}{\epsfig{file=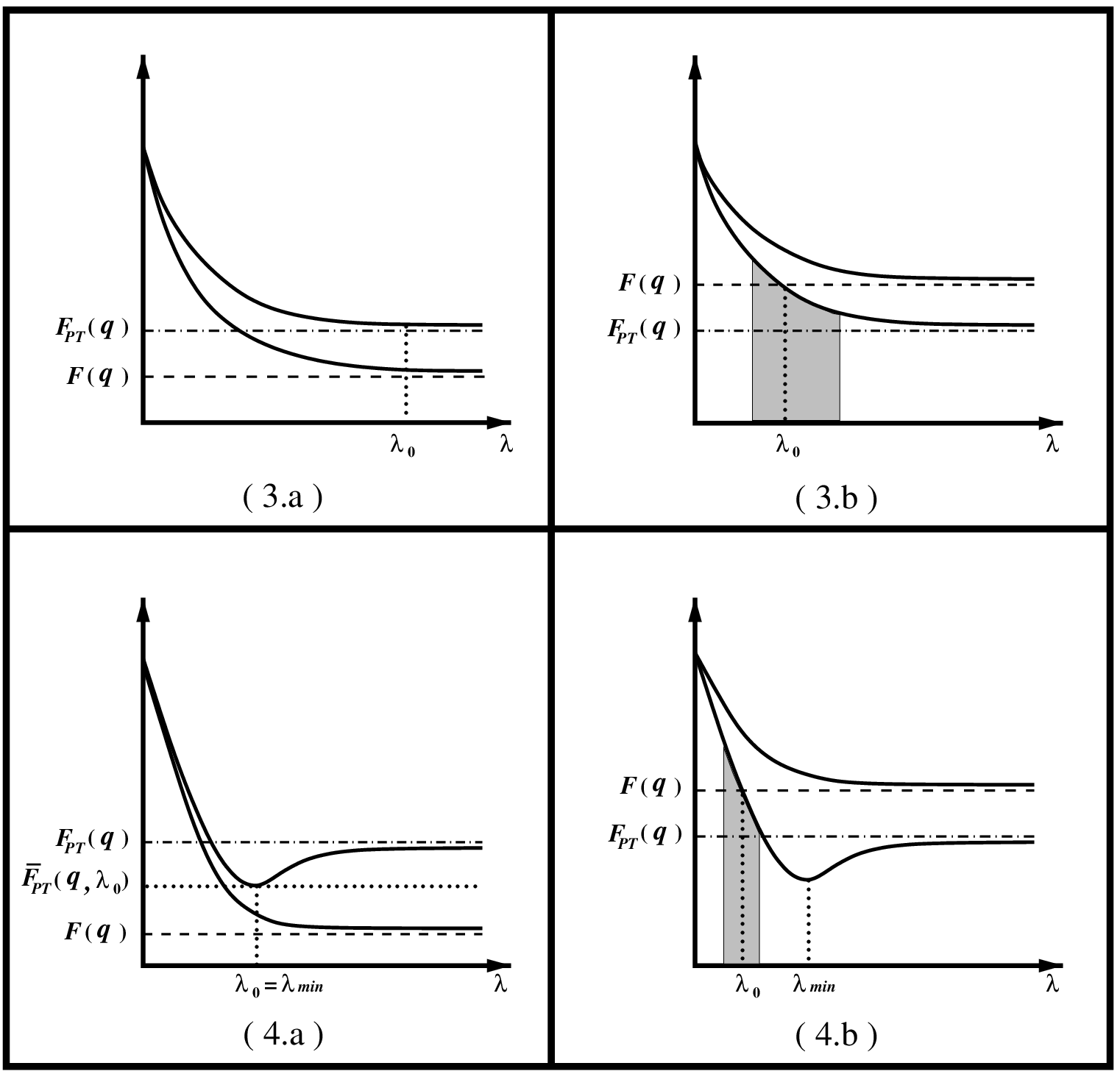}} 
\caption{The variation of $\bar F_{PT}(q,\lambda)$ and the expected 
$\bar F(q,\lambda)$ versus $\lambda$ at a fixed value $q>\Lambda^2$.
In (3.a) and (4.a) the upper and the lower curves describe $\bar 
F_{PT}(q,\lambda)$ and $\bar F(q,\lambda)$ respectively. In (3.b) 
and (4.b) the upper and the lower curves describe $\bar F(q,\lambda)$ 
and $\bar F_{PT}(q,\lambda)$ respectively. }
\label{Fig5} }

For each of these four cases, there corresponds two possibilities either 
\bq
\textrm{(a)}&&\hspace{0.1in} F(q)<F_{PT}(q)\qquad \textrm{or}\,, 
\nonumber \\ 
\textrm{(b)}&&\hspace{0.1in} F(q)>F_{PT}(q)\,. \nonumber
\eq
Accordingly, we shall denote the case associated with either possibility by 
case($i$.a) or case($i$.b) with $i=1,2,3\; \textrm{or}\; 4$. These cases are 
illustrated in Fig.(\ref{Fig4}) and Fig.(\ref{Fig5}). In these
figures, we include for each case an illustrative hypothetical graph
for $\bar{F}(q,\lambda)$ based on our knowledge of the small and large 
behaviour of $\bar F(q,\lambda)$, i.e. $\bar F(q,\lambda)\cong
\bar{F}_{PT}(q,\lambda)$ for sufficiently small $\lambda$ and 
$\bar F(q,\lambda)\cong F(q)$ for large enough $\lambda$. 
In applying our method we will need to know {\it a priori}
whether $F_{PT}(q)$ is less than or greater than $F(q)$, either by 
input from an experiment or by a knowledge of the signs of higher 
order corrections to the perturbation theory in a regime where 
this can be trusted.

Our choice of $\lambda_0$ for the different cases, at a fixed value 
$q>\Lambda^2$, is as follows:
\begin{enumerate}
\item
in case(1.a), the ideal value of $\lambda_0$ is the one 
at which $\bar F_{PT}(q,\lambda)$ intersects with $F(q)$ as shown 
in Fig.(\ref{Fig4}.1.a); such a point is not precisely known
although it may be estimated from a knowledge of higher order corrections
to $F_{PT}$ if these are available. We
expect it to occur near the point of greatest curvature of $\bar
F_{PT}(q,\lambda)$  as illustrated by the shaded region in
Fig.(\ref{Fig4}.1.a). So for this case we might
alternatively take $\lambda_0$ to be
the value of $\lambda$ where the curvature is greatest.
\item
in case(1.b), the monotonicity of
$\bar F_{PT}(q,\lambda)$ means that sampling this curve at any finite value
of $\lambda$ will fail to improve the perturbative estimate. However the use of
a large, but finite value will depart little from the perturbative estimate
(as the asymptotic behaviour is approached rapidly) and will have the merit
of reinstating causality. So our best choice for $\lambda_0$ is
the smallest $\lambda-$value at which $\bar F_{PT}(q,\lambda)$ is acceptably
close to $F_{PT}(q)$.
\item
in case(2.a), the ideal value of $\lambda_0$ is the one 
at which $\bar F_{PT}(q,\lambda)$ intersects with $F(q)$ as shown 
in Fig.(\ref{Fig4}.2.a); such a point is expected to lie within a narrow 
domain, on the left of $\lambda_{max}$, corresponding to a small arc 
segment just below $F_{PT}(q)$ as illustrated by the shaded region in 
Fig.(\ref{Fig4}.2.a); such a point might be estimated from a knowledge of
higher order perturbative corrections; however without further information 
it is not possible to see how to identify this region from a knowledge 
of the perturbative function $\bar F_{PT}(q,\lambda)$ alone.
\item
in case(2.b), the best choice for $\lambda_0$ is
$\lambda_{max}$, as illustrated in Fig.(\ref{Fig4}.2.b),  
and this leads to an improvement over the perturbative estimate. 
\end{enumerate}
\vspace{0.2in}
The cases depicted in Fig.(\ref{Fig5}) differ essentially by just a sign 
change, so a similar discussion of the appropriate choice of $\lambda_0$
can be made.

To summarise, from a {\it prior} knowledge of whether the perturbative
estimate overshoots or undershoots the true value of the observable 
$F(q)$, and by plotting the shape of $\bar F_{PT}(q,\lambda)$ against 
$\lambda$ we can ascertain which of the various possible cases occurs 
in a given situation. 
For cases (1.a), (2.a), (3.b) and (4.b), there exists a unique
point $\lambda=\lambda_0$ at which $\bar F_{PT}(q,\lambda_0)$ matches the 
observable $F(q)$. However, for cases (2.b) and (4.a) the situation is
different but still there is a point $\lambda_0$ at which $\bar F_{PT}(q,
\lambda_0)$ can give a bitter result than that of perturbation theory.  
In all of these cases, we can estimate $\lambda_0$ from the properties of 
$\bar F_{PT}(q,\lambda)$, or with the aid of higher order perturbative 
corrections, in a way that will improve on the perturbative predictions. 
On the other hand, in the remaining two cases (1.b) and (3.a) such 
an improvement does not occur.

Having determined $\lambda_0$ as a function of $q$ in a domain where
perturbation theory is reasonably trustworthy, we use its average 
$\lambda_e$ in $\bar F_{PT}(q,\lambda_e)$ to extrapolate to
smaller values of $q$. This results in an estimate free of the
unphysical divergences of the original perturbative approximation,
and this is the central goal of our approach. This will be true even 
for the cases (1.b) and (3.a).
In some cases, the effective average value $\lambda_e$ is either equal to 
or not significantly different from $\lambda_0(q)$. This is always true if 
$\lambda_0(q)$ is either a constant or a slowly varying function over a large 
domain of $q$. For instance, in our illustrative example $\lambda_{e}$ assumes
the value $\lambda_{max}$ in case(2.b) and $\lambda_{min}$ in case(4.a); in 
other words it coincides with $\lambda_{0}(q)$ for all $q>\Lambda^2$. Also, 
it follows from the inequality:
\be
|\Upsilon_{IR}(q,\lambda_e)| \le \,
\frac{|U(\Lambda^2)|}{q-\Lambda^2}\;\;
{\mathrm e}^{-\lambda_{e}(q-\Lambda^2)}+
\frac{|\omega|}{\pi \Lambda^2}\: 
\frac{{\mathrm e}^{-\lambda_e q}}{\lambda_e\, q } 
\qquad \textrm{for}\;\:q \ne \Lambda^2 \,,
\label{2.35}
\ee 
that the contribution of $\Upsilon_{IR}(q,\lambda_e)$ becomes negligible
for $q\gg \Lambda^2 $. Hence, employing $\lambda_e$ in the remote UV
region does not spoil the standard perturbative results.

One final point to remark is that the method also applies in the reverse 
direction to the previous setting. For example, in the case when the infrared 
behaviour $ F_{IR}(q) $ is well known then a similar representation to
(\ref{2.1}), but with $q$ replaced by a variable with the dimensions
of length squared $r=1/q $, can be used in almost the same way as 
before except this time to deduce the corresponding analytic UV expression:
\be 
\bar F_{UV}(q=1/r,\lambda)=\frac{1}{2\pi i}\, \int_{C_{IR}}
\, \,\frac{{\mathrm e}^{\lambda(\bar r-r)}}{\bar{r}-r}\,\,
F_{IR}(1/\bar r) \; d\bar r\,.
\label{2.36}
\ee
From this viewpoint, our method plays the role of a bridge between regions of 
small and large momenta, allowing us from a {\it prior} knowledge of either 
the UV or IR behaviour to extract information about the IR or UV properties 
respectively.

\section{Analysis of the one-loop QCD running coupling }

In this section, we demonstrate the implementation of our method in 
constructing a simple analytic model for describing the regular IR
behaviour of the QCD effective coupling constant. For ease of
calculation, we assume that the zeros of the exact function
$\alpha(Q^2)$ of the running coupling are expected to occur
only on the negative real axis $\mathrm{Re}\{Q^2\}<0$ and/or infinity.
This, together with the causality principle, allows us to express the 
reciprocal of $\alpha(Q^2)$ in terms of the contour integral:
\be 
\frac{1}{\alpha(Q^2)}=\frac{1}{2\pi i}\: \lim_{\lambda \rightarrow \infty}
\:\;\int_{C_{UV}}\, \,\frac{{\mathrm e}^{\lambda(k-Q^2)}}{k-Q^2}\:\:
\frac{1}{\alpha(k)}\:\; dk\,,  
\label{3.1}
\ee 
introduced in the previous section. When applying this representation 
to an n-loop perturbative expression $\alpha^{(n)}_{\mathrm{PT}}(Q^2)$, 
we take $\lambda=\lambda_e$ instead of $\infty$. In this way, we obtain 
a combination of perturbative and non-perturbative contributions, which 
has the merit of:
\begin{enumerate}
\item
reinstating the causal analyticity structure  
in the right half plane $\mathrm{Re}\{Q^2\}\!\ge \!0$,
\item
preserving the standard UV behaviour and,  
\item
improving the perturbative estimate in both
the IR and low UV regions.  
\end{enumerate}
\vspace{0.1in}\noindent 
In the following, we shall prove this claim within the one-loop
approximation. For convenience, we express the reciprocal of 
$\alpha^{(1)}_{\mathrm{PT}}(Q^2)$ in terms of a dimensionless variable 
$q=Q^2/\Lambda^2$ as:
\be 
\frac{1}{\alpha^{(1)}(q)}=\frac{\beta_0}{4\pi }\,\ln(q)\,,  
\label{3.2}
\ee
where $\alpha^{(1)}(Q^2/\Lambda^2)=\alpha^{(1)}_{\mathrm{PT}}(Q^2)$
\footnote{In this paper, we use  the notation $\alpha^{(n)}(q=Q^2/\Lambda^2)=
\alpha^{(n)}_{\mathrm{PT}}(Q^2)$ for the corresponding n-loop approximation.}.
In the spirit of (\ref{3.1}), an analytically improved expression
corresponding to the one-loop coupling constant can be defined as:
\be
\bar \alpha^{(1)}(q,\lambda_e)=
\frac{1}{\bar \chi^{(1)}(q,\lambda)}\,\Bigg |_{\lambda=\lambda_e}\,,
\label{3.3}
\ee 
with
\be 
\bar \chi^{(1)}(q,\lambda)=\frac{1}{2\pi i}\,\int_{C_{UV}}
\, \,\frac{{\mathrm e}^{\lambda(k-q)}}{k-q}\;\: 
\frac{1}{\alpha^{(1)}(k)}\; dk\,. 
\label{3.4}
\ee
Note that $[\alpha^{(1)}(k)]^{-1}$, being dependent on $\ln q$,
has a cut along the negative real axis $\textrm{Re}\{ k\} \le 0$. 
We shall now proceed to evaluate the above
integral for the case $q\neq 0$. Following the same argument of the 
preceding section from (\ref{2.1}) to (\ref{2.4}), we can show that:
\be 
\bar \chi^{(1)}(q,\lambda)=\frac{\beta_0}{4\pi }\,\ln(q)+
\frac{\beta_0}{4\pi}\,I_{C_k}(q,\lambda)\,,
\label{3.5}
\ee
where
\be
I_{C_k}(q,\lambda)=\mathrm{J}_{\epsilon}(q,\lambda)+
\int^{R_{\scriptscriptstyle{\mathrm{UV}}}}_{\epsilon}
\;\frac{{\mathrm e}^{-\lambda(k+q)}}{k+q}\;dk \,,
\label{3.6}
\ee
with
\be
\mathrm{J}_{\epsilon}(q,\lambda)=
\frac{\epsilon\,{\mathrm e}^{-\lambda q}}{2 \pi}\;
\int^{\pi}_{-\pi}\,{\mathrm e}^{\lambda\epsilon\cos(\theta)}\;\;
\textrm{Re}\left(\frac{{\mathrm e}^{i\lambda\epsilon\sin(\theta)}\,
\ln(\epsilon\; {\mathrm e}^{i\theta})}
{\epsilon-q\,{\mathrm e}^{-i\theta}}\right)\,d\theta\,.
\label{3.7}
\ee
A simple way to check that the contribution of
$\mathrm{J}_{\epsilon}(q,\lambda)$ vanishes as $\epsilon$ approaches zero 
is to consider the upper bound:
\be
|\mathrm{J}_{\epsilon}(q,\lambda)|\leq \;
\frac{{\mathrm e}^{-\lambda(q-\epsilon)}}{q-\epsilon}\;\;
\epsilon \,\left(\ln^2(\epsilon)+\pi^2\right)^{1/2}\,,
\label{3.8}
\ee
from which it follows that:
\be
\lim_{\epsilon \rightarrow 0}\;\mathrm{J}_{\epsilon}(q,\lambda)=0\,.
\label{3.9}
\ee
Consequently, in the limit as $\epsilon\to 0$ we can write in a simple form:
\be
I_{C_k}(q,\lambda)=\,\int^{R_{\mathrm{UV}}}_{0} \,
\frac{{\mathrm e}^{-\lambda(k+q)}}{(k+q)} \,dk\;=
E_1(\lambda q)-E_1(\lambda(q+R_{\mathrm{\scriptscriptstyle{UV}}}))\,. 
\label{3.10}
\ee
By inserting this into (\ref{3.5}) and letting $R_{\mathrm
{\scriptscriptstyle{UV}}}\to \infty$, we obtain:
\be 
\bar \chi^{(1)}(q,\lambda)=\frac{\beta_0}{4\pi }\,\Big[\,\ln(q)+
E_1(\lambda\, q)\,\Big]\,.
\label{3.11}
\ee

Having found the exact structure of $\bar \chi^{(1)}(q,\lambda)$ 
for $q\ne 0$, let us now reconsider the contour integral (\ref{3.4}) 
once more to explore the continuity of $\bar \chi^{(1)}(q,\lambda)$ 
at the origin $q=0$. Starting from the expression:
\be 
\bar \chi^{(1)}(0,\lambda)=\frac{\beta_0}{4\pi}\frac{1}{2\pi i}\,\int_{C_{UV}}
\, \,\frac{{\mathrm e}^{\lambda k}}{k}\,\ln(k) \, dk\,,  
\label{3.12}
\ee
we arrive, after some work, at
\begin{align}
\bar \chi^{(1)}(0,\lambda)=& \frac{\beta_0}{4\pi}\, 
\lim_{\epsilon \rightarrow 0}
\, \Bigg[\; \int_{\epsilon}^{R_{\mathrm{UV}}}\,\,
\frac{{\mathrm e}^{-\lambda k}}{k} \, dk 
\, + \, \frac{\ln(\epsilon)}{\pi}\,\int_{0}^{\pi}\,{\mathrm e}^{\lambda 
\epsilon\cos(\theta)}
\;\cos(\lambda \epsilon \sin(\theta)) \; d\theta  
\nonumber\\
&\qquad\quad -\frac{1}{\pi} \, \int_{0}^{\pi}\,{\mathrm e}^{\lambda \epsilon 
\cos(\theta)}
\; \sin(\lambda \epsilon \sin(\theta))\,\theta \; d\theta \:\, \Bigg]\,.
\label{3.13}
\end{align}
Making use of
\be
\int_{\epsilon}^{ R_{\mathrm{UV}}} \,\,\frac{{\mathrm e}^{-\lambda k}}{k} \, 
dk\,=
E_1(\lambda \epsilon)-E_1(\lambda R_{\scriptscriptstyle{\mathrm{UV}}})\,,
\label{3.14}
\ee
and inserting the following series representations \cite{Gradshteyn80}:
\begin{align}
{\mathrm e}^{\lambda \epsilon\cos(\theta)}\;\cos(\lambda \epsilon
\sin(\theta))=&
\sum_{n=0}^{\infty} \;\frac{(\lambda \epsilon)^n}{n!}\;\cos(n \theta) 
\qquad\textrm{for}\;\;\lambda \epsilon <1\,,
\nonumber\\
{\mathrm e}^{\lambda \epsilon\cos(\theta)}\;\sin(\lambda \epsilon 
\sin(\theta))=&
\sum_{n=1}^{\infty} \;\frac{(\lambda \epsilon)^n}{n!}\;\sin(n \theta) 
\qquad \textrm{for}\;\;\lambda \epsilon <1\,,\nonumber
\end{align}
into the two angular integrals in (\ref{3.13}), we immediately obtain:
\be
\bar \chi^{(1)}(0,\lambda) = \frac{\beta_0}{4\pi}\, 
\lim_{\epsilon \rightarrow 0}\, 
\Bigg[\;E_1(\lambda \epsilon)-E_1(\lambda
R_{\mathrm{\scriptscriptstyle{UV}}})
+\,\ln(\epsilon)+\sum_{n=1}^{\infty}\;\frac{(-1)^n}{n! \, n}\;
(\lambda \epsilon)^n \:\,\Bigg]\,.
\label{3.15}
\ee
If we now make use of (\ref{2.18}) and let 
$R_{\mathrm{\scriptscriptstyle{UV}}}\to \infty$, we arrive at: 
\be
\bar \chi^{(1)}(0,\lambda) =\frac{\beta_0}{4\pi}\,
\ln({\mathrm e}^{-\gamma}/\lambda)\,.
\label{3.16}
\ee
This ensures the continuity condition at the origin:
\be
\lim_{q\rightarrow 0}\;\bar \chi^{(1)}(q,\lambda) =
\bar \chi^{(1)}(0,\lambda)\,.
\label{3.17}
\ee

Let us now find the range of $\lambda$ for which:
\be
\bar \alpha^{(1)}(q,\lambda)=
\frac{1}{\bar\chi^{(1)}(q,\lambda)}=
\frac{4\pi }{\beta_0}\,
\frac{1}{[\;\ln(q)+E_1(\lambda\, q)\;]} \,,
\label{3.18}
\ee
is continuous on the whole interval $q\ge 0$. This requires 
an investigation of the zeros of $\bar \chi^{(1)}(q,\lambda)$ 
in $D_{\bar\chi}$, where $D_{\bar\chi}=\{(q,\lambda):q\ge 0,\;
\lambda >0\}$. From expression (\ref{3.11}), we know that since 
$E_1(\lambda\,q)>0$ in $D_{\bar\chi}$ and $\ln q$ is negative
only when $q\in(0,1)$, the zeros of the function 
$\bar \chi^{(1)}(q,\lambda)$ are expected to occur somewhere 
in the subinterval\break 
$\{(q,\lambda):0\le q<1,\;\lambda >0\}\subset D_{\bar\chi}$.
This domain can be reduced further by first considering 
the complementary exponential integral:
\be
Ein(x)=\int_0^1
\frac{[1-\mathrm{e}^{-xt}]}{t}\;dt
=\sum_{n=1}^{\infty}\,\frac{(-1)^{n-1}}{n}\;
\frac{x^n}{n!}\,,\qquad x\ge 0\,,
\label{3.18a}
\ee
to rewrite (\ref{3.11}) as:
\be
\bar\chi^{(1)}(q,\lambda)=
\frac{\beta_0}{4\pi}\,
[\;\ln( \mathrm{e}^{-\gamma }/\lambda )
+Ein(\lambda\,q)\;] \,.
\label{3.18b}
\ee
Then by observing the fact that $Ein(\lambda\,q)\ge 0$ for all
$\lambda\,q\ge 0$ and that $\ln(\mathrm{e}^{-\gamma}/\lambda)<0$ 
only if $\lambda>\mathrm{e}^{-\gamma }$, we deduce that the 
zeros of $\bar\chi^{(1)}(q,\lambda)$ can exist only in the following 
interval $\{(q,\lambda):0\le q<1,\;\lambda\ge
\mathrm{e}^{-\gamma}\}$. 
Hence, $\bar \alpha^{(1)}(q,\lambda)$ is continuous for all 
$q\ge 0$ if and only if $\lambda\in(0,\mathrm{e}^{-\gamma})$. 
Another way to obtain this result is by considering the fact 
that $\bar\alpha^{(1)}(0,\lambda)\ge
\bar\alpha^{(1)}(q,\lambda)$ for all values of $\lambda$ 
that keep $\bar\alpha^{(1)}(q,\lambda)$ finite and positive 
throughout the range $q\ge 0$. Then, from the positivity of:
\FIGURE[ht]{ \unitlength 1cm
\put(1,11.5){$\lambda$} \put(12.5,1.05){$q$}
\put(7.3,3.9){$\mathrm{e}^{-\gamma}\approx\,\scriptstyle{0.561}$}
\resizebox{13cm}{12cm}{\epsfig{file=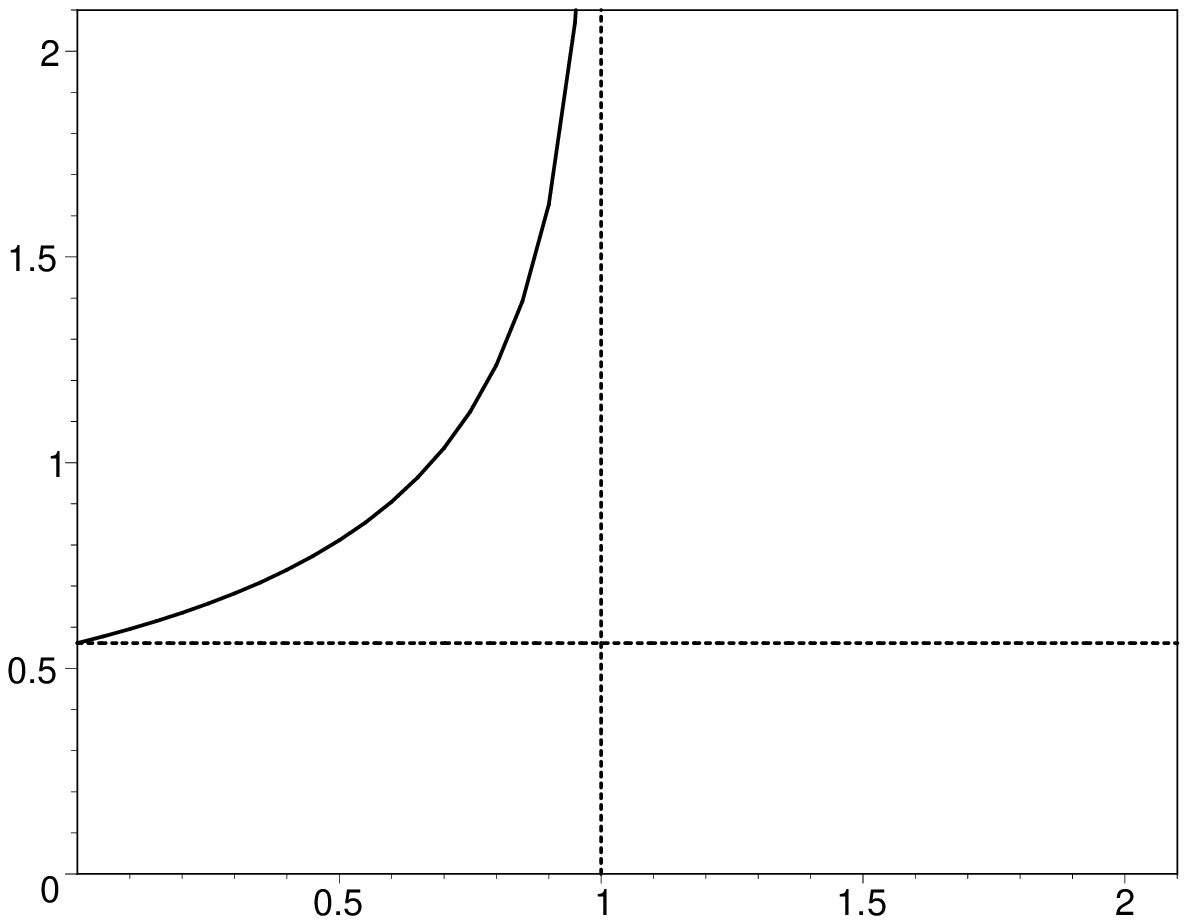}}
\caption{The zeros of $\bar \chi^{(1)}(q,\lambda)$:
the curve shown in the strip $\{ (q,\lambda):0\le q<1,\;
\lambda \ge e^{-\gamma}\}$ represents the roots of 
$\bar \chi^{(1)}(q,\lambda)$ in $D_{\bar\chi}$. 
This curve never touches the vertical $q=1$ as 
$\bar \chi^{(1)}(q,\lambda)$ does not have a zero 
on this line.}
\label{Figch2.1} }
\be
\bar \alpha^{(1)}(0,\lambda)=\frac{4\pi }{\beta_0} \; 
\frac{1}{\ln({\mathrm e}^{-\gamma}/\lambda)}\,,
\label{3.19}
\ee
it follows that the allowed values of $\lambda$ are those 
confined to the interval:
\be
D_{eff}=\{\lambda\,:\;0<\lambda<{\mathrm e}^{-\gamma}
\approx 0.561\}\,.
\label{3.20}
\ee
Using Maple VI for solving $\bar \chi^{(1)}(q,\lambda)=0$ 
numerically, we illustrate in Fig.(\ref{Figch2.1}) the zeros 
of $\bar\chi^{(1)}(q,\lambda)$, i.e the singularities of 
$\bar \alpha^{(1)}(q,\lambda)$, by a curve confined to the 
strip $\{(q,\lambda):0\le q<1,\;\lambda \ge e^{-\gamma}\}$,
showing that $D_{eff}$ is the only region of $\lambda$ for which 
$\bar \alpha^{(1)}(q,\lambda)$ is continuous for all $q\ge 0$.

\section{The Estimation of $\lambda_e$}

In this section, we try to find an effective value $\lambda_e
\in D_{eff}$ such that $\bar\alpha^{(1)}(q,\lambda_e)$ provides 
a better estimate than that of the standard perturbative results,
especially in the vicinity of the Landau-pole and the IR region.

From the constraint (\ref{3.20}), i.e. $\lambda_e\in D_{eff}$,  
it follows that:
\be
0<\bar\alpha^{(1)}(q,\lambda_e)<\textrm{M(q)} \qquad
\textrm{for all}\;\;q\ge 0\,, 
\label{3.21}
\ee 
where the upper bound 
$\textrm{M(q)}=\bar\alpha^{(1)}(q,{\mathrm e}^{-\gamma})$.
This implies that 
$\bar\alpha^{(1)}(q,\lambda_e)<\alpha^{(1)}(q)=
\bar\alpha^{(1)}(q,\lambda\gg 1)$ for all $q>1$.
However, since $E_1(\lambda_e\,q)$ in (\ref{3.18}) is negligibly 
small for $q\gg \lambda_e $, we may consider 
$\bar\alpha^{(1)}(q,\lambda_e)=\alpha^{(1)}(q)$ at 
sufficiently large values of $q$.
For a given $n_f$ and $q>1$, we find that the way in which 
$\bar\alpha^{(1)}(q,\lambda)$ changes with $\lambda$, as shown in
Fig.(\ref{Fig6}), is a typical example of case(1) illustrated 
in Fig.(\ref{Fig4}).
\FIGURE[htb]{ \unitlength 1cm 
\put(6.5,0.5){$\lambda$}\put(14.1,0.5){$\lambda$}
\put(3.5,-0.3){(a)}\put(11.2,-0.3){(b)}
\put(8,5.6){\small{$\bar\alpha^{(1)}(2,\lambda)$}}
\put(5,5.27){\small{$\alpha^{(1)}(2)$}}
\put(5.1,4){\small{$\textrm{M(2)}$}}
\put(1,3){\small{$\bar\alpha^{(1)}(2,\lambda)$}}
\resizebox{7.1cm}{6.1cm}{\epsfig{file=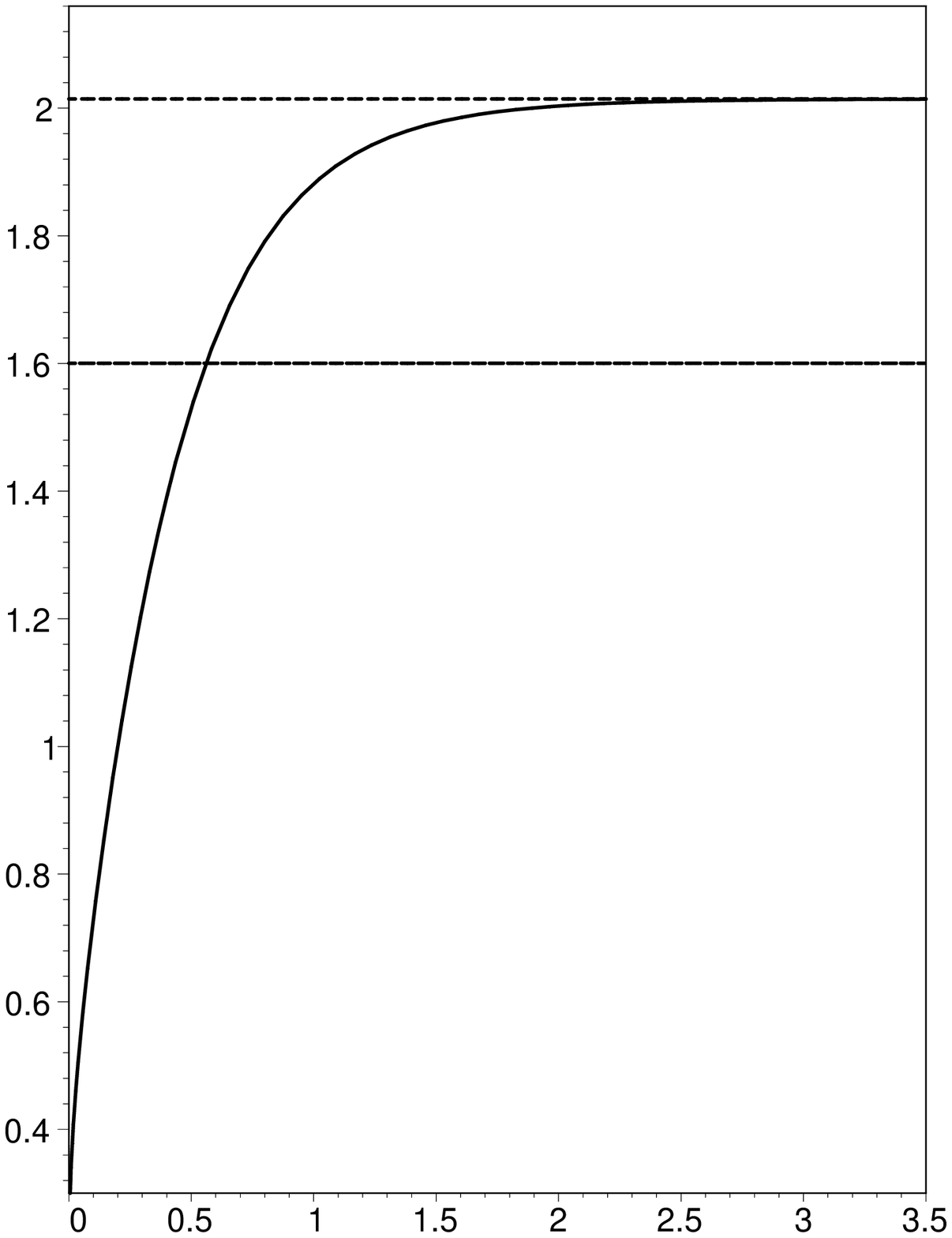}}
\hspace{0.1cm}
\resizebox{7.1cm}{6.1cm}{\epsfig{file=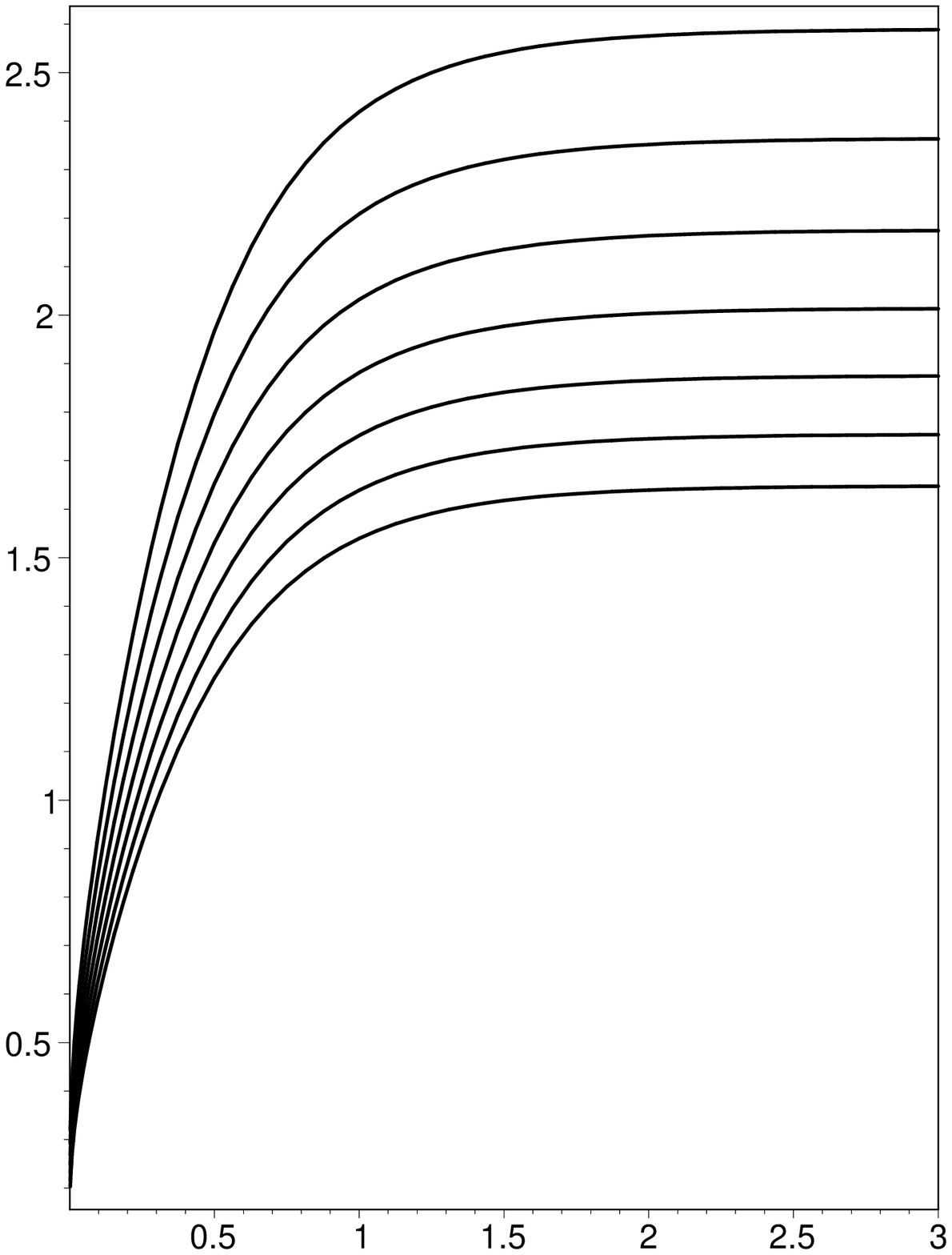}}
\caption{The variation of $\bar \alpha^{(1)}(2,\lambda)$ versus 
$\lambda$, (a) for $n_f=3$ and (b) for $0 \le n_f \le 6$. 
\break Here, $\bar\alpha^{(1)}(2,\lambda)$ tends to $\alpha^{(1)}(2)$
as $\lambda$ increases.
In (b), $\bar\alpha^{(1)}(2,\lambda)$ increases 
with $n_f$ for any fixed value of $\lambda$. }
\label{Fig6} }
Hence, before setting up a criterion for
determining $\lambda_e$, we need to find out whether the
perturbative estimate $\alpha^{(1)}(q)$, in the low UV region, overshoots 
or undershoots the true value $\alpha(q)$. This can be deduced by
comparing $\alpha^{(1)}(q)$ to the standard higher-loop corrections
\cite{Yndurain99}:
\be
\alpha^{(2)}(q)=\frac{4\pi}{\beta_0\,L}\:\Big[1-\frac{\beta_1}{\beta_0^2}
\:\frac{\ln L}{L}\Big]\label{3.22}\,,
\ee
and 
\be
\alpha^{(3)}(q)=\frac{4\pi}{\beta_0\,L}\:\Big[1-\frac{\beta_1}{\beta_0^2}
\:\frac{\ln L}{L}+\frac{\beta_1^2\,\ln^2 L-\beta_1^2\,\ln L+
\beta_2\,\beta_0-\beta_1^2}{\beta_0^4\;L^2}\Big] \label{3.23}\,,
\ee
where $L=\ln q\,$ with $q=Q^2/\Lambda^2$, and:
\bq
\beta_1 &=& 102-\frac{38}{3}\,n_f\,,\label{A3.24}\\
\beta_2 &=& \frac{2857}{2}-\frac{5033}{18}\,n_f+\frac{325}{54}\,n_f^2\,.
\label{A3.25}
\eq 
Fig.(\ref{Fig7}) illustrates the comparison in terms of the physical scale 
$Q$ [GeV].
\FIGURE[ht]{ \unitlength 1cm 
\put(5.2,-0.3){$Q$ \small{[GeV]}}
\put(3.1,4.6){\small{$\alpha^{(1)}_{\mathrm{PT}}(Q^2)$}}
\put(2.6,2.86){\small{$\alpha^{(2)}_{\mathrm{PT}}(Q^2)$}}
\put(1.85,1.59){\small{$\alpha^{(3)}_{\mathrm{PT}}(Q^2)$}}
\resizebox{11cm}{9cm}{\epsfig{file=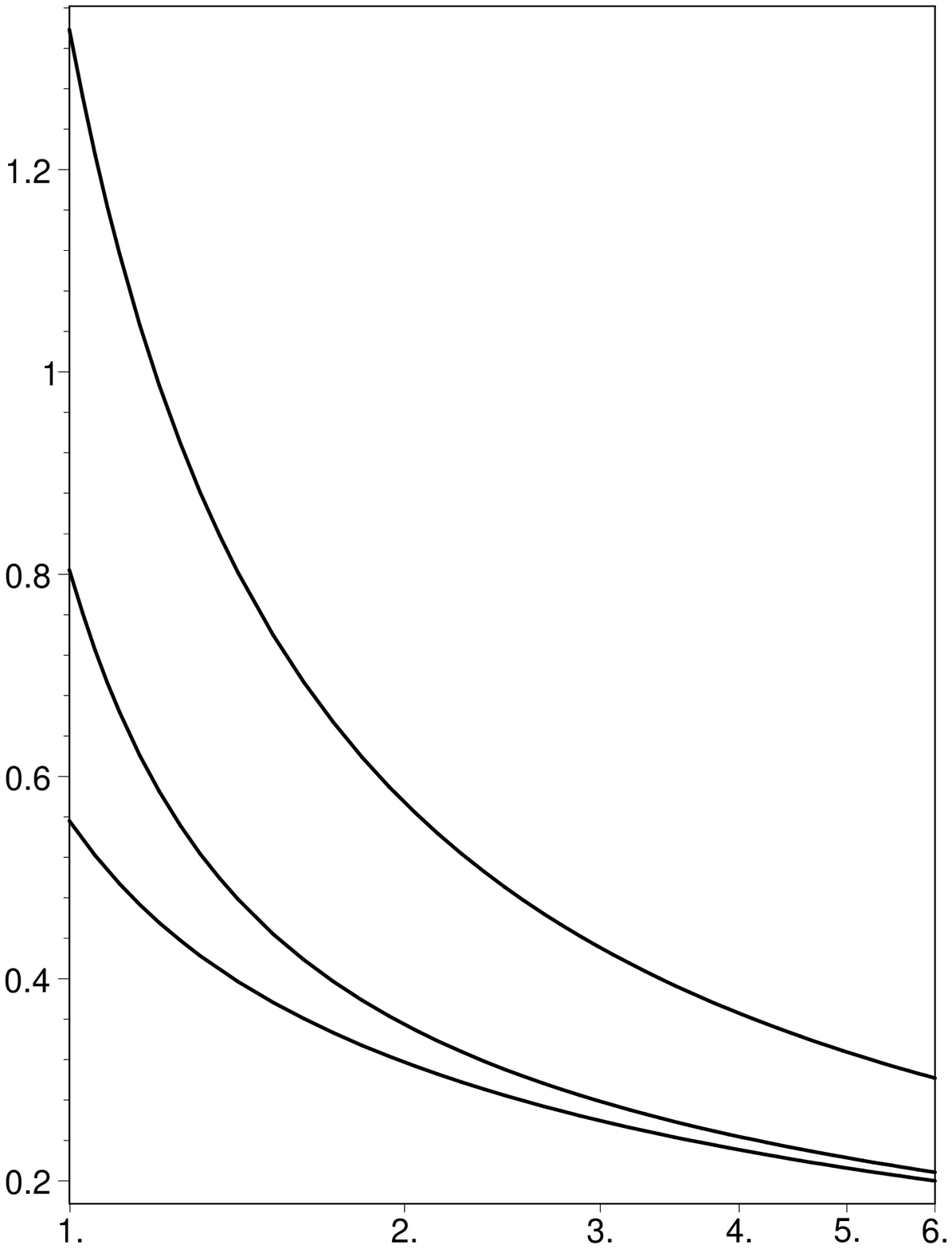}}
\caption{The comparison between the 1-, 2- and 3-loop running 
coupling for $n_f=3$. }
\label{Fig7} }
In this figure, we consider $n_f=3$ and take $\Lambda =$ 0.593 GeV, 
0.495 GeV and 0.383 GeV for $\alpha^{(1)}_{\mathrm{PT}}(Q^2)$, 
$\alpha^{(2)}_{\mathrm{PT}}(Q^2)$ and $\alpha^{(3)}_{\mathrm{PT}}(Q^2)$ 
respectively. The $\Lambda$ values are determined in the 
$\overline{\textrm{MS}}$ scheme from the experimentally measured 
$\tau-$lepton decay rate \cite{Korner01} given by:
\be
R_{\tau}=2.9087\:(0.998+\delta_p^{(n)})=3.492 \label{3.24}\,,
\ee
where
\be 
\delta_p^{(n)}=\sum_{k=1}^{n}a_k\:\Bigg(\frac{\alpha^{(n)}_{\mathrm{PT}}
(M_{\tau}^2)}{\pi}\Bigg)^k\label{3.25}\,,
\ee
here $n\le 3$ denotes the $n^{th}$ loop order approximation, $a_1=1$,
$a_2=5.20232$, $a_3=26.3659$ and $M_{\tau}=1.777$ GeV.
In Fig.(\ref{Fig7}), we used $n_f=3$ as the average number of active
quarks, ignoring complications due to quark thresholds. This is
reasonable in the low energy interval $Q<3$ GeV. For a more 
precise description of the evolution of $\alpha^{(n)}_{\mathrm{PT}}(Q^2)$
in a larger momentum interval, one should take into account the effects of
quark thresholds. However, this does not change the overall picture
illustrated in Fig.(\ref{Fig7}). 
Having now observed that $\alpha^{(1)}_{\mathrm{PT}}(Q^2)>
\alpha^{(2)}_{\mathrm{PT}}(Q^2)>\alpha^{(3)}_{\mathrm{PT}}(Q^2)$,
we would expect $\alpha^{(1)}(q)$ to overshoot the true value
of the coupling constant, at least for intermediate and sufficiently
small values of $q$.
This indicates that our problem is of the type classified earlier as
case(1.a) and depicted in Fig.(\ref{Fig4}.1.a). Hence, we may ascertain 
that for every point $q>1$ there exists a proper value $\lambda_0(q)$ such 
that $\bar \alpha^{(1)}(q,\lambda_0(q))=\alpha(q)$, just as explained in 
case(1.a). One way to estimate $\lambda_0(q)$ is to first consider
the curvature function of $\bar \alpha^{(1)}(q,\lambda)$:
\be 
K^{(1)}(\lambda,q)=\;\frac{|\bar\alpha^{(1)}_{\lambda\,\lambda}(q,\lambda)|}
{\Big[\;1+(\bar\alpha^{(1)}_{\lambda}(q,\lambda))^2\,\Big]^{3/2}}\,,
\label{3.26}
\ee
for a fixed $q>1$, where
\begin{align}
\bar\alpha^{(1)}_{\lambda}(q,\lambda)&=\frac{\partial}
{\partial\lambda}\bar \alpha^{(1)}(q,\lambda)=
\frac{\beta_0}{4\pi}\,\big[\,
\bar\alpha^{(1)}(q,\lambda)\,\big]^2\;
\frac{\mathrm{e}^{-\lambda q}}{\lambda}\,,
\label{3.26a}\\
\bar\alpha^{(1)}_{\lambda\,\lambda}(q,\lambda)&
=\frac{\partial^2}
{\partial\lambda^2}\bar \alpha^{(1)}(q,\lambda)=
\bar\alpha^{(1)}_{\lambda}(q,\lambda)\,
\Big[\,\frac{\beta_0}{2\pi}\;
\bar\alpha^{(1)}(q,\lambda)\;
\frac{\mathrm{e}^{-\lambda q}}{\lambda}
-\frac{1}{\lambda}-q\,\Big]\,.
\label{3.26b}
\end{align}
Then take $\lambda_0(q)$ to be the value of $\lambda$ at 
which $K^{(1)}(\lambda_0,q)$ is a local maximum. 
For sufficiently small values of $q>1$, $K^{(1)}(\lambda,q)$ 
has two local maxima, say at $\lambda_1$ and $\lambda_2$. 
In this case, we consider $\lambda_0$ to be the largest of these 
(say $\lambda_0=\lambda_2>\lambda_1$) so that 
$\bar \alpha^{(1)}(q,\lambda_0)$ remains as close as
possible to the perturbative results $\alpha^{(1)}(q)$, 
which are still reasonable for such values of $q$. 
In Fig.(\ref{Fig8}) we plot the curvature function 
$K^{(1)}(\lambda,q)$ versus $\lambda$ for some fixed values
of $q$, illustrating in part (a) of the figure the difference
between $\bar \alpha^{(1)}(q,\lambda_0)$ and $\alpha^{(1)}(q)$ 
for $q=3$. In part (b) of Fig.(\ref{Fig8}), we observe that the 
value of $\lambda_0$ decreases with increasing $q$.
\FIGURE[ht]{ \unitlength 1cm 
\put(6.8,0.5){$\lambda$}\put(14.4,0.5){$\lambda$}
\put(3.5,-0.3){(a)}\put(11.1,-0.3){(b)}
\put(4.5,4.8){\small{$\bar\alpha^{(1)}(3,\lambda)$}}
\put(4.5,3.58){\small{$\bar\alpha^{(1)}(3,\lambda_0)$}}
\put(3.4,1.6){\small{$K^{(1)}(\lambda,3)$}}
\put(1.8,0.5){$\lambda_0$}
\put(11.48,1.4){\small{$K^{(1)}(\lambda,3)$}}
\put(9.7,3.7){\small{$K^{(1)}(\lambda,4)$}}
\put(9.2,5){\small{$K^{(1)}(\lambda,5)$}}
\resizebox{7.2cm}{6.1cm}{\epsfig{file=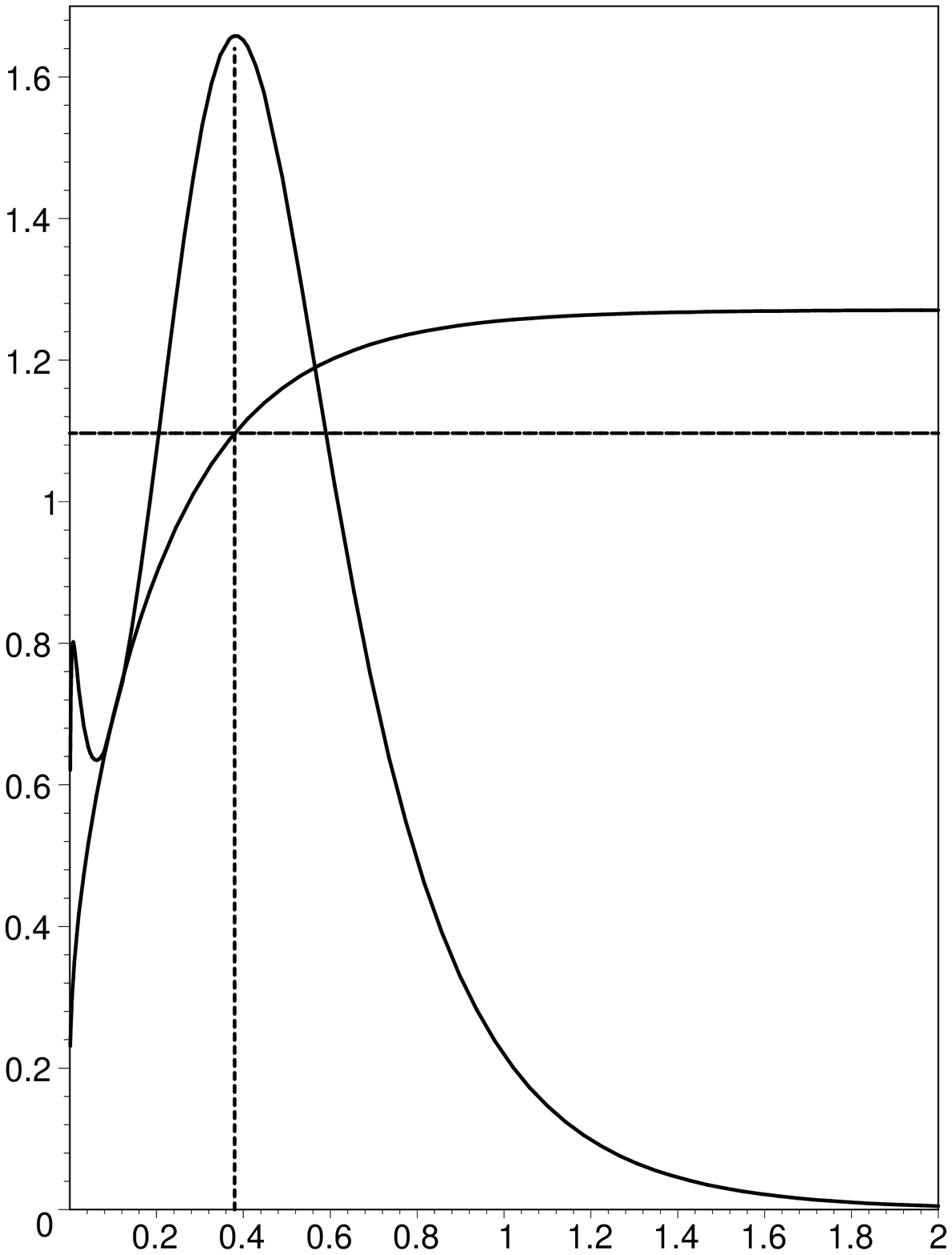}}
\hspace{0.1cm}
\resizebox{7.2cm}{6.1cm}{\epsfig{file=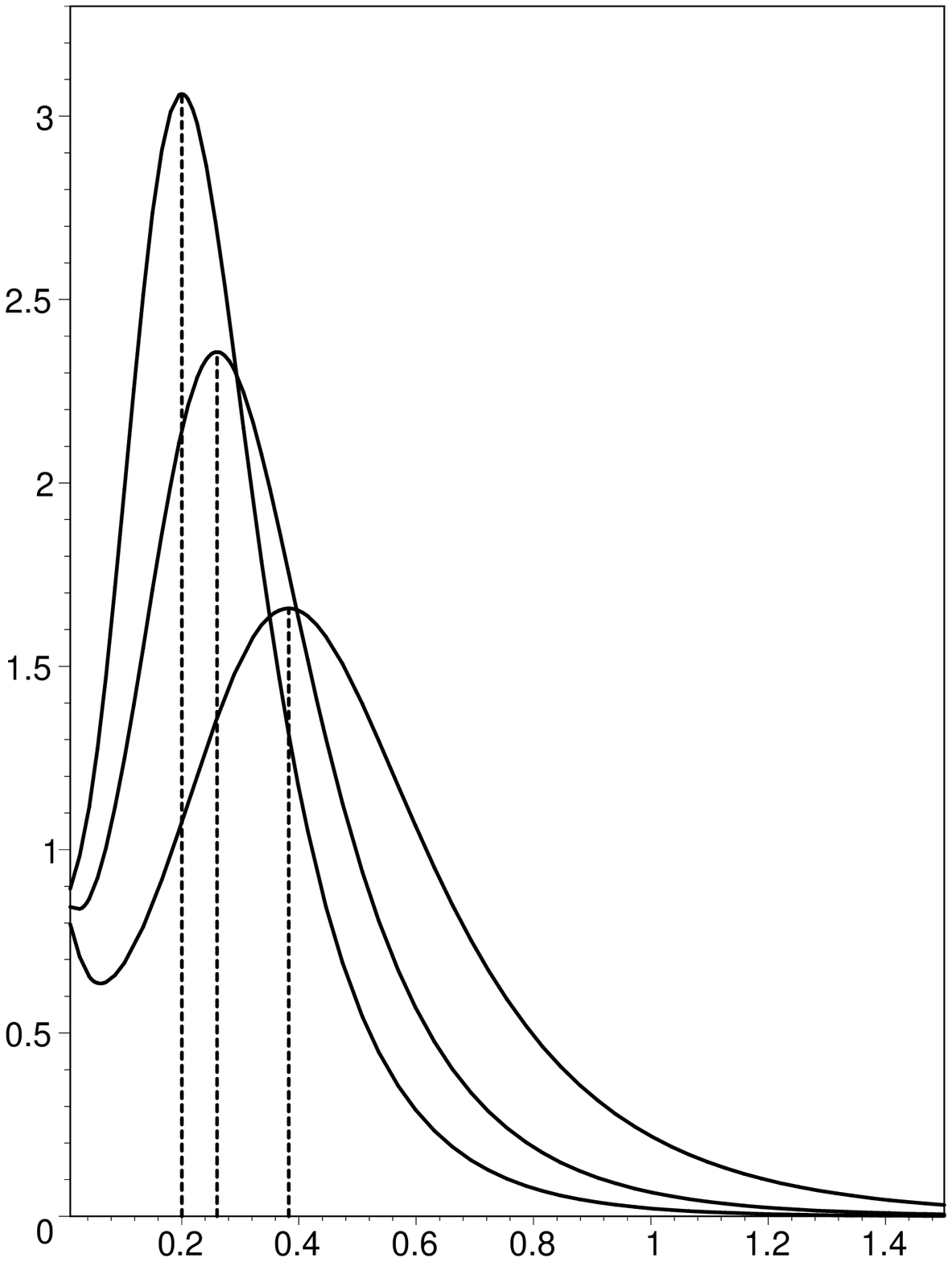}}
\caption{The variation of $K^{(1)}(\lambda;q)$ versus $\lambda$ 
for $n_f=3$, (a) at $q=3$ and (b) at $q=3,4,5$. For a fixed $q>1$, 
$\,\lambda_0(q)$ denotes the value of $\lambda$ at which
$K^{(1)}(\lambda;q)$ is a local maximum.\break  
In (b), $\lambda_0(5)<\lambda_0(4)<\lambda_0(3)$.}
\label{Fig8} }
In general, if we plot $\lambda_0$ against $q$ for any quark flavour 
$n_f$ we find that $\lambda_0(q)$ is a monotonically decreasing 
function as shown in Fig.(\ref{Fig9}) for $n_f=3$.
\FIGURE[ht]{ \unitlength 1cm 
\put(7.5,0.52){q}\put(7.5,2.8){$\lambda_e$}
\put(1.65,5){\small{$\lambda_0(q)$}}
\resizebox{14.5cm}{6.5cm}{\epsfig{file=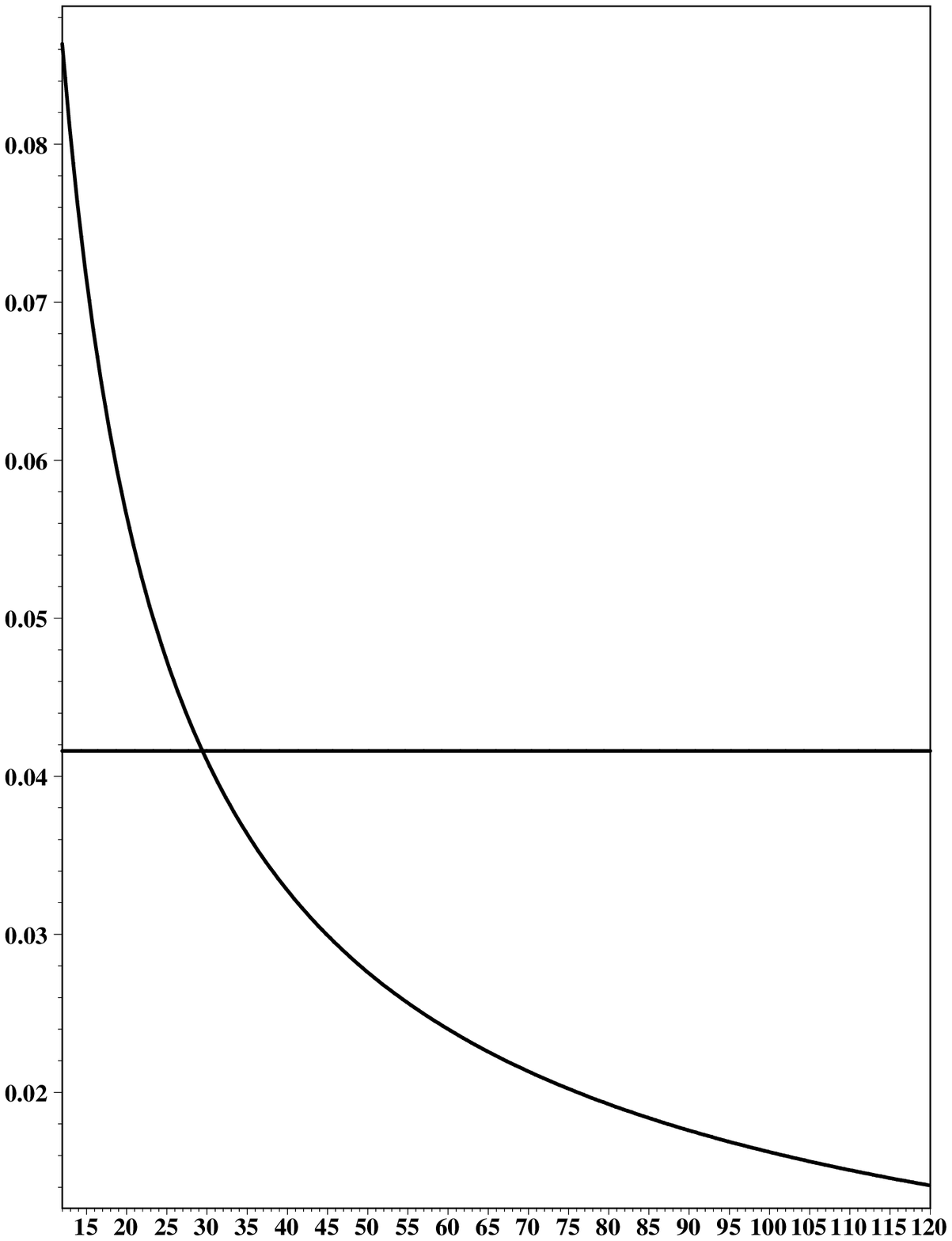}}
\caption{The variation of $\lambda_0(q)$ versus $q$, for $n_f=3$.}
\label{Fig9} }
Hence, we may express $D_{eff}$ in terms of $\lambda_0(q)$ as:
$$
D_{eff}=\{\lambda_e:\lambda_0(\infty)<\lambda_e<\lambda_0(q_0)\}\,,
$$
with $q_0$ defined by $\lambda_0(q_0)={\mathrm e}^{-\gamma}$. 
The values of $q_0$ for different $n_f$ are all given in
table(\ref{T1}). \noindent
\TABLE[ht]{\centerline{
\begin{tabular}{|c|c|c|c|c|c|c|c|}
\hline
$n_f$  &   0   &    1  &   2   &   3   &   4   &  5    &   6  \\ \hline
$q_0$  & 2.204 & 2.246 & 2.293 & 2.345 & 2.402 & 2.467 & 2.539 \\ \hline
\end{tabular}}
\caption{Numerical values of $q_0(n_f)$}
\label{T1}}     
For sufficiently large values of $q$, our choice of $\lambda_0$ does 
not lead to a significant improvement on the perturbative estimate. 
In fact, since only the first two decimal places of the coupling 
constant are of physical interest, we may take 
$\bar\alpha^{(1)}(q,\lambda_0(q))\cong\alpha^{(1)}(q)$ 
for all $q\ge120$ and $0 \le n_f\le 6 $. On this basis, we shall 
focus our attention on the effective domain:
\be
S_{eff}=\{\;q:\;q_0<q<120\;\}
\label{3.27}\,,
\ee
over which the difference between $\bar\alpha^{(1)}(q,\lambda_0(q))$ 
and $\alpha^{(1)}(q)$ is appreciable. Thus, we expect to find 
$\lambda_e$ in a narrower interval, $\bar D_{eff}=
\{\lambda_e:\,\lambda_0(120)<\lambda_e<\lambda_0(q_0)\}
\subset\,D_{eff}\,$. Since, as illustrated in Fig.(\ref{Fig9}), 
$\lambda_0(q)$ is a very slowly varying function over a large 
part of the interval $S_{eff}$, i.e. for $q\in (11,120)$, 
we take the average of $\lambda_0(q)$ over $S_{eff}$ to be 
the best estimated value for $\lambda_e$. 
Our numerical results for $\lambda_e$ are listed in 
table(\ref{T2}) for the $n_f$ values of practical interest. 
As seen from table(\ref{T2}), the values of $\lambda_e$ do not
change appreciably with $n_f$. Hence, we shall fix $\lambda_e$ 
to a central value of 0.04 for all $n_f$.     
\TABLE[ht]{\centerline{
\begin{tabular}{|c|c|c|c|c|c|c|c|}
\hline
\small
$n_f$       &    0  &    1  &   2  &   3  &   4   &    5  &   6   \\ \hline
$\lambda_e$ &0.0379 &0.0390 &0.0402&0.0416&0.0429 &0.0444 & 0.0461 \\ \hline
\end{tabular} }
\caption{Numerical values of $\lambda_e(n_f)$}
\label{T2} }

Let us now denote $\bar \alpha^{(1)}(Q^2/\Lambda^2,\lambda_e)$ by 
$\bar \alpha^{(1)}(Q^2)$, where:
\be
\bar \alpha^{(1)}(Q^2)=\frac{4\pi }{\beta_0}\,
\frac{1}{[\;\ln(Q^2/\Lambda^2)+E_1(\lambda_e\,Q^2/\Lambda^2)\;]} \,.
\label{3.28}
\ee
In this expression, the exponential integral function 
$E_1(\lambda_e\,Q^2/\Lambda^2)$ plays an important role in 
removing the ghost pole at $Q=\Lambda$, preserving the correct UV 
behaviour (as it decays rapidly to zero for $Q\gg\Lambda$) and enforcing 
the running coupling to freeze in the low IR region. Therefore, it is
essentially non-perturbative. At low energies, $\bar \alpha^{(1)}(Q^2)$ 
approximates well to the simple formula:
\be
\bar \alpha^{(1)}(Q^2)=\frac{4\pi}{\beta_0\,\kappa_e}\,\;
\frac{\varpi^2}{\varpi^2+Q^2}\qquad 
\textrm{for}\quad 0\le Q^2\le\Lambda^2 \,,
\label{3.29}
\ee
where $\kappa_e=\ln({\mathrm e}^{-\gamma}/\lambda_e)$ and 
$\varpi^2=\kappa_e\;\Lambda^2/\lambda_e$.
In particular, this yields a finite value at the origin, namely:
\be
\bar \alpha^{(1)}(0)=\frac{4\pi}{\beta_0\,\kappa_e}\,,
\label{3.30}
\ee
which is totally independent of the characteristic mass scale $\Lambda$.
Obviously, this is a key advantage of our approach as it agrees with the 
IR freezing phenomenological hypothesis \cite{Mattingly94,Stevenson94}.
Now, let me remark from Ref.\cite{Stevenson94} that the freezing 
phenomenon is not incompatible 
with confinement as there is no evidence that confinement necessarily 
requires the coupling constant to become infinite in the IR region. 
For instance, Gribov's ideas \cite{Stevenson94,Close93} explicitly involve 
a freezing of the coupling constant at low momenta. This phenomenon is 
also present in perturbation theory but beyond the leading order and for 
certain numbers of quark flavours.


\section{Empirical Investigations of Our Model}
In this section, we compare our approach with other recognised methods
including conventional and modified perturbation theory. Then, we test
our model on phenomenologically estimated data for a fit-invariant 
characteristic integral depending solely on the IR behaviour of the 
coupling constant.

\subsection{Threshold Matching}
In this subsection, we shall discuss the method we use for determining 
the values of $\Lambda^{(n_f)}$ to be employed in our calculation. 
For sufficiently large values of $Q^2$, our expression for the coupling
constant reduces to the familiar one-loop formula $\alpha^{(1)}_{\mathrm{PT}}
(Q^2)$. This suggests that our $\Lambda^{(n_f)}$ must coincide with the usual
QCD scale parameter $\Lambda^{(n_f)}_{\mathrm{PT}}$ in momentum intervals 
corresponding to $n_f\ge 5$. Hence, we set our 5-flavour mass parameter 
$\Lambda^{(5)}$ equal to $\Lambda^{(5)}_{\mathrm{PT}}$ at the energy scale of
the Z-boson mass $\mathrm{M}_Z=91.19$ GeV. We extract the value $\Lambda^{(5)}
=0.135$ GeV from $\alpha^{(1)}_{\mathrm{PT}}(\mathrm{M}_Z^2)$ by means of the
recent parametrisation of the hadronic decay width of the Z-boson 
\cite{Tournefier98}:
\be
R_Z=19.934\;\Bigg [1+\sum_{k=1}^{n}a_k\:\Bigg(\frac{\alpha^{(n)}_{\mathrm{PT}}
(\mathrm{M}_Z^2)}{\pi}\Bigg)^k\;\Bigg ]  \label{4.1}\,,
\ee
where $n\le 3$ denotes the $n^{th}$ loop order under consideration, $a_1=1.045$,
$a_2=0.94$, $a_3=-15$ and $R_Z=20.768$ \cite{LEP}. 
To obtain the values of $\Lambda^{(4)}$ and $\Lambda^{(3)}$, which may
differ from $\Lambda^{(4)}_{\mathrm{PT}}$ and $\Lambda^{(3)}_
{\mathrm{PT}}$ as they correspond to lower energy intervals, we use
the familiar matching condition \cite{Marciano84}:
\be
\bar \alpha^{(1)}(Q_{th};n_f)=\bar \alpha^{(1)}(Q_{th};n_f-1)\,,
\label{4.2}
\ee
at the energy thresholds $Q_{th}=2\,m_b$ and $Q_{th}=2\,m_c$, 
where $m_b=4.21$ GeV and $m_c=1.35$ GeV are the bottom- and 
charm-quark masses $\cite{Mattingly94}$ respectively. 
In this way, starting with $\Lambda^{(5)}=0.135$ GeV, 
we get $\Lambda^{(4)}=0.188$ GeV and $\Lambda^{(3)}=0.229$ GeV.

At light quark thresholds $Q_{th}\le 2\,m_s$, where $m_s=0.199$ GeV is
the strange-quark mass $\cite{Mattingly94}$, 
the applied matching condition (\ref{4.2}) holds only 
approximately. This is also true for the analytic perturbation theory of
Shirkov and Solovtsov \cite{Shirkov97}. For simplicity, instead of introducing
a nontrivial and more complicated matching procedure such as the one used
in Ref.\cite{Chetyrkin00}, we choose to take $\Lambda^{(n_f)}=\Lambda^{(3)}$
for all $n_f\le 3$. This choice is quite reasonable since it leads to:
\be
\frac{\bar \alpha^{(1)}(Q_{th};n_f)}{\bar \alpha^{(1)}(Q_{th};n_f-1)}\cong 
1\,,
\label{4.3}
\ee
for all energy thresholds corresponding to $n_f<3$.
Furthermore, it is also supported by the good agreement with 
the value $\Lambda^{(0)}=0.238\,(0.019)$ GeV obtained from quenched 
lattice QCD in Ref. \cite{Capitani99}.

\subsection{Illustrative comparison}

We begin by considering a quick comparison between the predictions in our 
approach and those in perturbation theory at one-, two- and three-loop 
order.
\FIGURE[ht]{ \unitlength 1cm 
\put(5.3,-0.23){\small{$Q$ [GeV]}}\put(13.8,-0.23){\small{$Q$ [GeV]}}
\put(3.5,-0.7){(a)}\put(11.4,-0.7){(b)}
\put(9,4.8){$\scriptstyle{\alpha}^{\scriptscriptstyle{(1)}}_
{\mathrm{\scriptscriptstyle{PT}}}$}
\put(8.38,3.9){$\scriptstyle{\alpha}^{\scriptscriptstyle{(2)}}_
{\mathrm{\scriptscriptstyle{PT}}}$}
\put(7.92,3.19){$\scriptstyle{\alpha}^{\scriptscriptstyle{(3)}}_
{\mathrm{\scriptscriptstyle{PT}}}$}
\put(8.1,1.26){$\bar{\scriptstyle{\alpha}}^{\scriptscriptstyle{(1)}}$}
\put(3,5.3){$\bar\alpha^{(1)}(Q^2)$}
\centerline{\resizebox{7cm}{7cm}{\epsfig{file=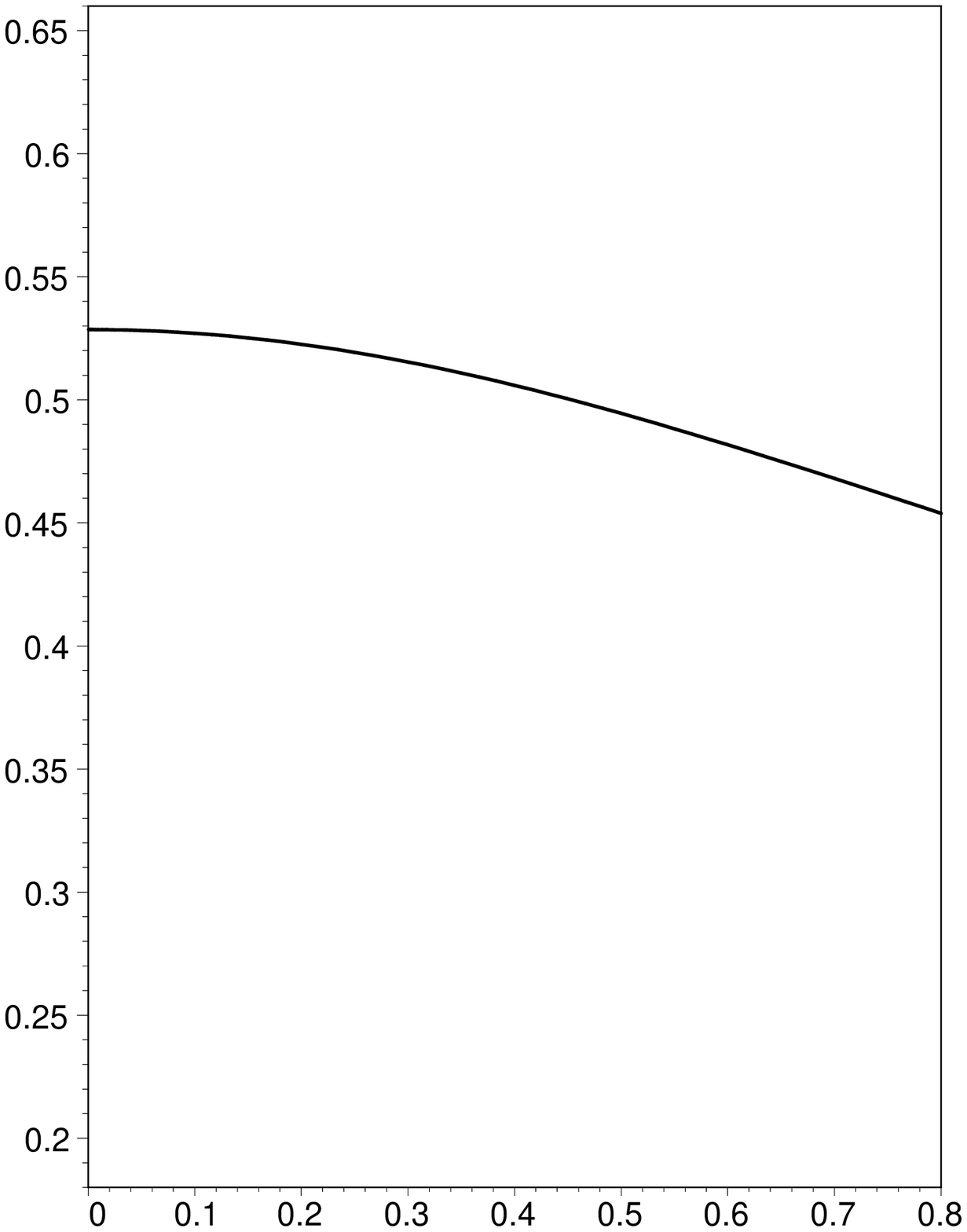}}
\hspace{0.1cm}
\resizebox{8cm}{7cm}{\epsfig{file=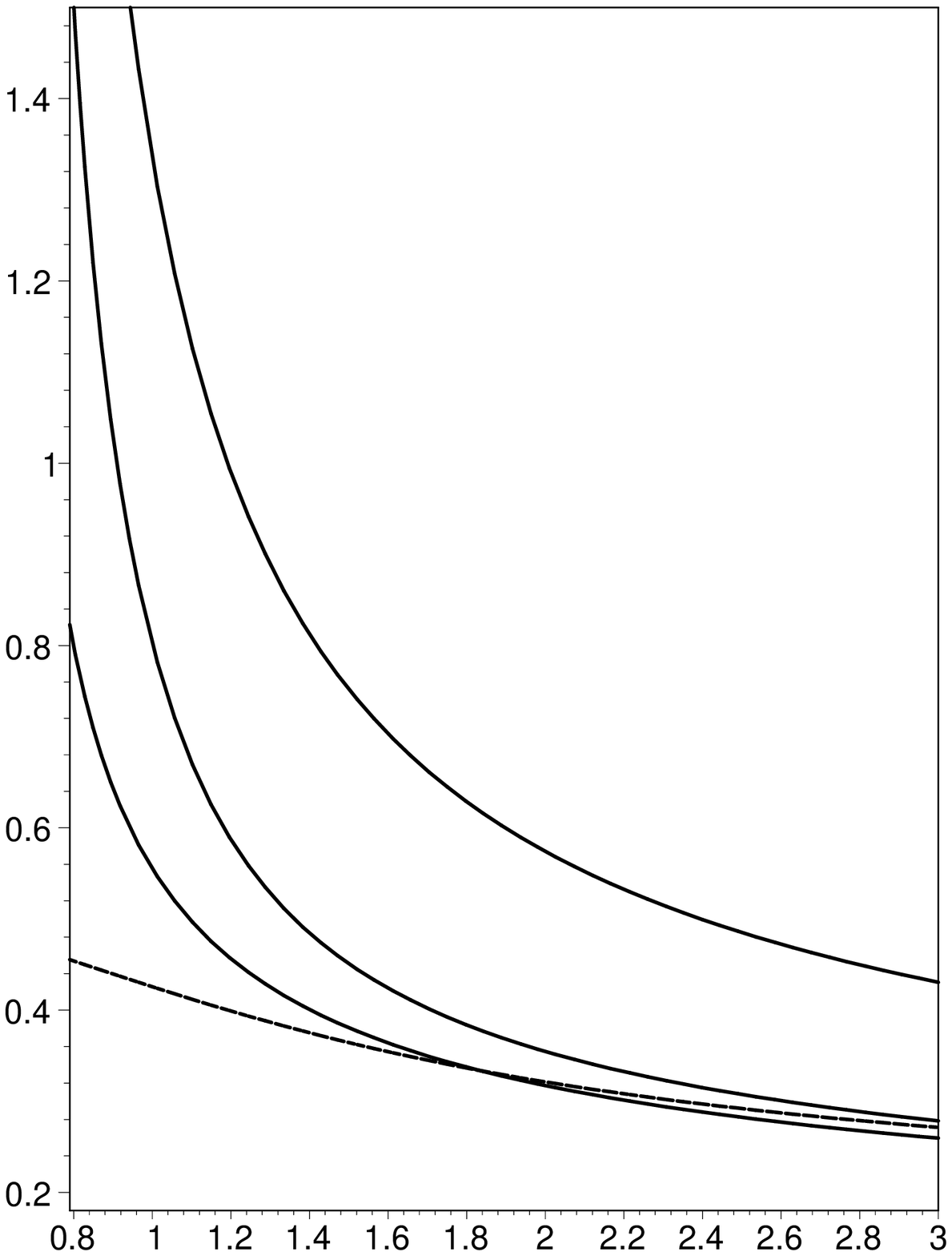}}}
\caption{The IR freezing behaviour of our model (a) and a comparison of
our model with the 1-, 2- and 3-loop perturbative expressions (b). }
\label{Fig10} }
Fig.(\ref{Fig10}) shows the behaviour of the 3-flavour running coupling 
$\bar\alpha^{(1)}(Q^2)$ together with $\alpha^{(1)}_{\mathrm{PT}}(Q^2)$, 
$\alpha^{(2)}_{\mathrm{PT}}(Q^2)$ and $\alpha^{(3)}_{\mathrm{PT}}(Q^2)$ 
in the low UV region as well as the Q-dependence of $\bar\alpha^{(1)}(Q^2)$ 
outside the perturbative domain. In this plot we use the same $\Lambda$ 
values employed in Fig.(\ref{Fig7}) and take $\Lambda^{(3)}=0.229$ GeV 
for our formula $\bar \alpha^{(1)}(Q^2)$. As is seen from
Fig.(\ref{Fig10}.b), our model improves significantly on 
the 1-loop perturbative predictions and agrees well with the 
higher-loop estimates throughout the range $Q>1.5$ GeV. 
In Fig.(\ref{Fig10}.a), we observe that the evolution of 
$\bar\alpha^{(1)}(Q^2)$ slows down appreciably as $Q$ enters the IR region, 
$Q\le\Lambda^{(3)}$, and freezes rapidly to a finite value of 0.529. 
This provides some direct theoretical evidence for the freezing of the 
running coupling at low energies, an idea that has long been a popular 
and successful phenomenological hypothesis. Phenomenological studies, 
such as \cite{Mattingly94} and references therein, 
show that a running coupling that freezes at low energies can be useful 
in describing experimental data and some IR effects in QCD. 
In the low energy domain: $0.8\;\textrm{GeV} \le Q\le 1.2\;\textrm{GeV}\,,$
Badalian and Morgunov \cite{Badalian99} extracted the value of the strong 
coupling constant $\alpha_s(Q^2)$ from the fits to charmonium spectrum 
and fine structure splittings. They found that $\alpha_s(Q^2)=0.38
\pm0.03(exp.)\pm0.04(theory)$. This agrees  reasonably well with our 
prediction, $\bar\alpha^{(1)}(Q^2)=0.42\pm0.03$ in the same energy interval.

Within the 3-loop approximation of the optimized perturbation theory, 
Mattingly and Stevenson \cite{Mattingly94} found that the 2-flavour QCD 
running coupling $\alpha_s$ freezes below 0.3 GeV to a constant value 
of about 0.26 $\pi$. Although this value is quantitatively uncertain 
(while qualitatively unequivocal), it is not far off from our 1-loop 
prediction of 0.16 $\pi$.

A good theoretical approach which supports the IR freezing behaviour 
of our model follows from the work of Simonov and Badalian 
\cite{Simonov93,Badalian97}.
They studied the non-perturbative contribution to the QCD running 
coupling on the basis of a new background field formalism, exploiting
non-perturbative background correlation functions as a dynamical input. 
In this framework, the one- and two-loop running coupling constants 
\cite{Badalian97}:
\bq 
\alpha^{(1)}_B(Q^2)&=&\frac{4\,\pi}{\beta_0}\;\frac{1}{\ln[(Q^2+M^2_B)/
\Lambda_B^2]}\,,
\label{4.4}\\
\nonumber \\
\alpha^{(2)}_B(Q^2)&=&\frac{4\,\pi}{\beta_0}\;\frac{1}{\ln[(Q^2+M^2_B)/
\Lambda_B^2]}\;\Bigg[1-\frac{\beta_1}{\beta_0^2}\frac{\ln[
\ln[(Q^2+M^2_B)/\Lambda_B^2]]}         
{\ln[(Q^2+M^2_B)/\Lambda_B^2]}\Bigg]\,,
\label{4.5}
\eq                        
were found to depend on a new parameter $M_B$ (similar to $\lambda_e$ in our
model) known as the background mass. From a fit to the charmonium fine 
structure \cite{Badalian99}, it was found that $M_B=1.1$ GeV. 
In table(\ref{T3}), we list the values of $\Lambda_B^{(n_f)}$ 
for $n_f\le 5$ in leading order (LO) and next-to-leading order (NLO). 
They are obtained as described in Ref.\cite{Badalian97}, 
using $\Lambda_{\mathrm{PT}}^{(5)}=0.135$ GeV in LO and
$\Lambda_{\mathrm{PT}}^{(5)}=0.274$ GeV in NLO, which is basically the same
method that we used to obtain the values of our $\Lambda^{(n_f)}$.
\TABLE[ht]{\centerline{
\begin{tabular}{|c|c|c|c|c|c|c|c|}
\hline
\small
$          n_f      $    &  0   &  1   &  2   &    3  &   4  &  5   
\\ \hline
$\Lambda_B^{(n_f)}$ GeV  & 0.307& 0.283& 0.258& 0.230& 0.188& 0.135 
\\ LO  & & & & & &\\ \hline
$\Lambda_B^{(n_f)}$ GeV  & 0.569& 0.553& 0.535& 0.509& 0.414& 0.274 
\\ NLO & & & & & &\\ \hline
\end{tabular}}
\caption{The values of $\Lambda_B^{(n_f)}$ in LO and NLO }
\label{T3}}

In Fig.(\ref{Fig11}), we display the comparison between the 3-flavour 
\FIGURE[ht]{ \unitlength 1cm 
\put(5.9,-0.23){\small{$Q$ [GeV]}}\put(13.8,-0.23){\small{$Q$ [GeV]}}
\put(3.5,-0.7){(a)}\put(11.4,-0.7){(b)}
\put(9.5,4.6){$\scriptstyle{\alpha}^{\scriptscriptstyle{(2)}}_
{\scriptscriptstyle{B}}$}
\put(8.48,4.05){$\bar{\scriptstyle{\alpha}}^{\scriptscriptstyle{(1)}}$}
\put(9.5,2.3){$\scriptstyle{\alpha}^{\scriptscriptstyle{(1)}}_
{\scriptscriptstyle{B}}$}
\put(2.5,5.61){$\alpha^{(2)}_{B}(Q^2)$}
\put(2.5,2.55){$\alpha^{(1)}_{B}(Q^2)$}
\put(2.5,4){$\bar\alpha^{(1)}(Q^2)$}
\centerline{\resizebox{7.5cm}{7cm}{\epsfig{file=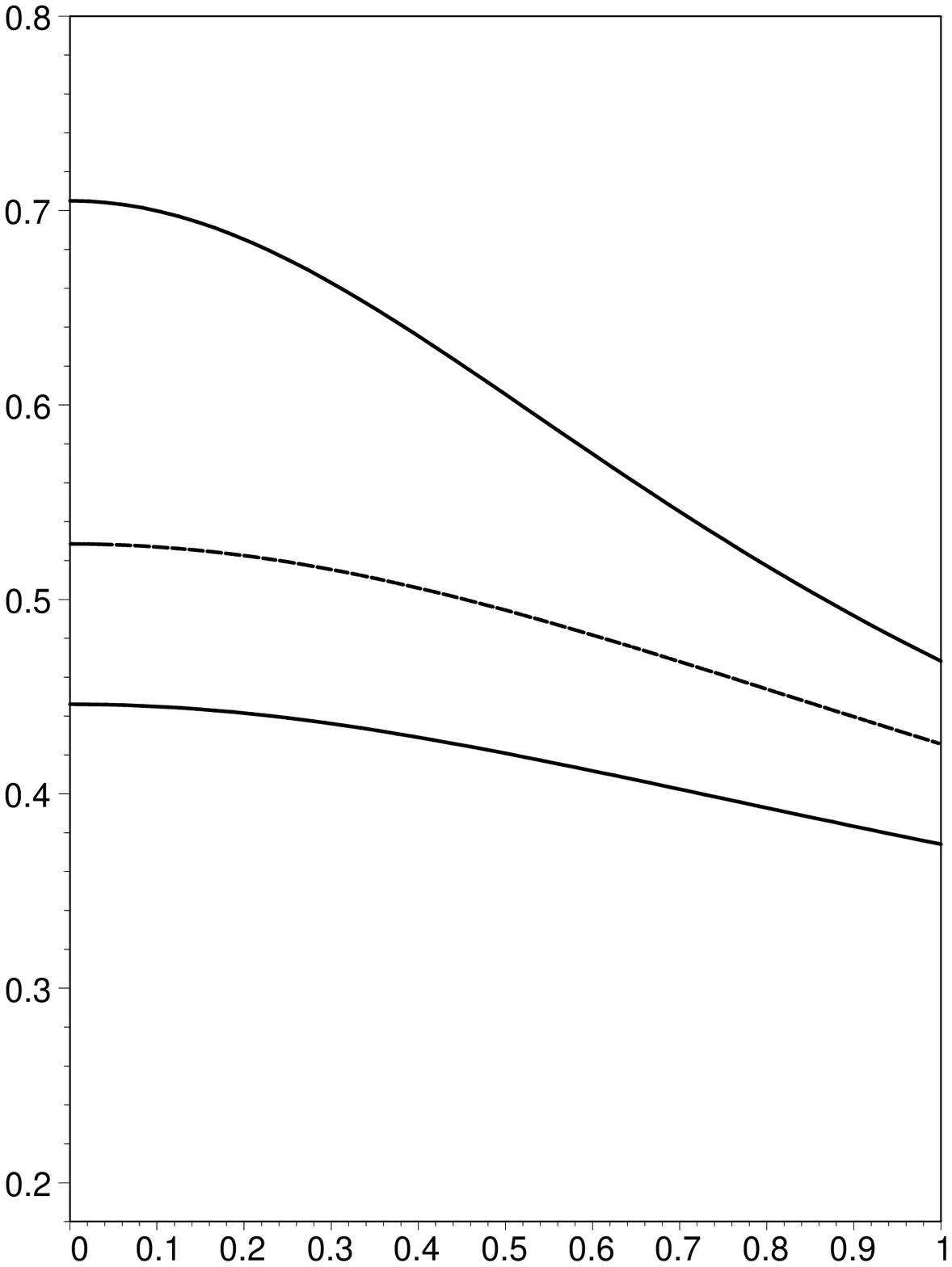}}
\hspace{0.1cm}
\resizebox{7.5cm}{7cm}{\epsfig{file=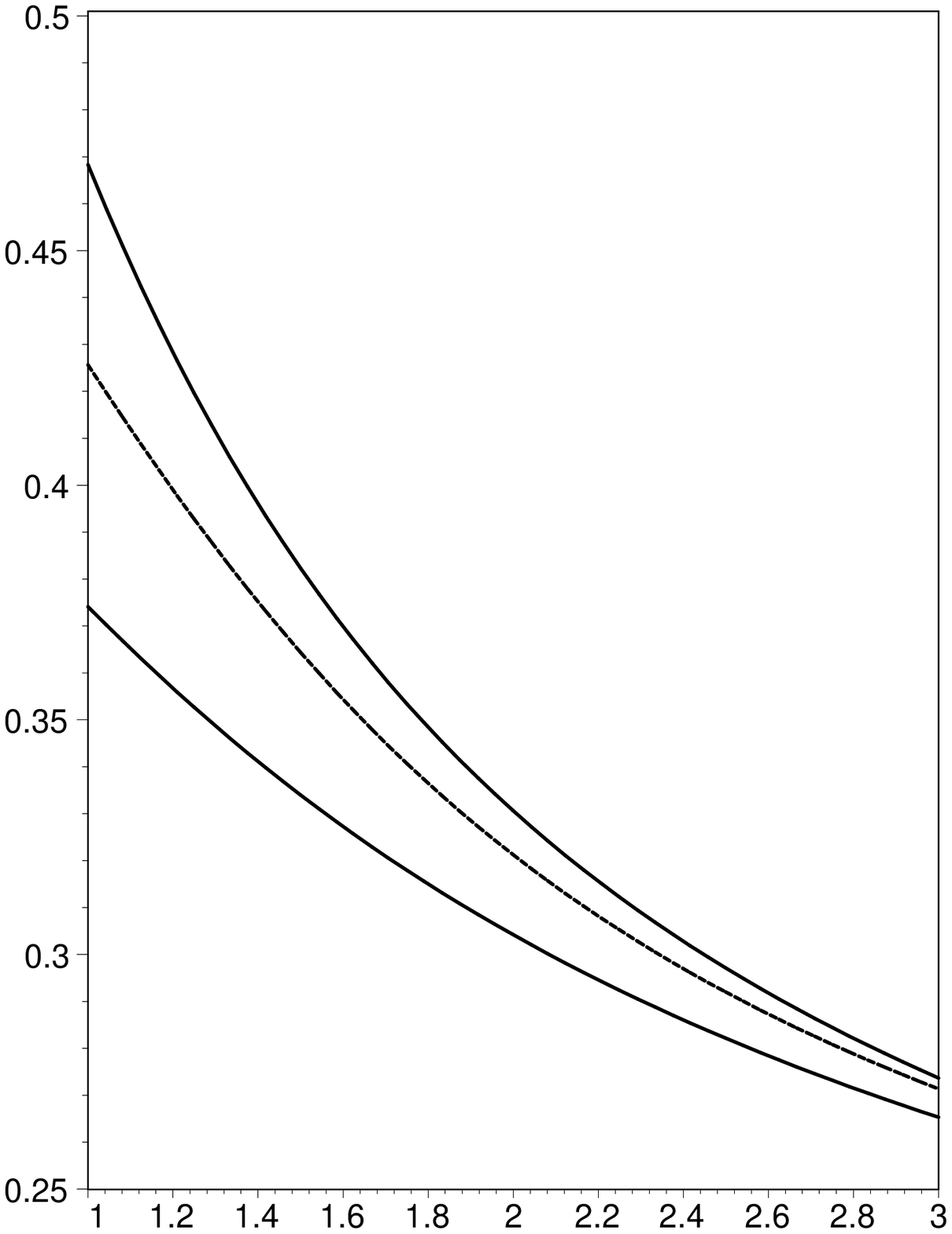}}}
\caption{Illustrative comparison of our formula $\bar\alpha^{(1)}(Q)$ 
with the LO and NLO expressions for the background coupling constants
$\alpha^{(1)}_B(Q)$ and $\alpha^{(2)}_B(Q)$ for $n_f=3$\,.}
\label{Fig11} }
coupling constant in our approach and those in the new background 
field formalism calculated to leading and next-to-leading order. 
As shown in the figure, our low energy estimates lie approximately 
in the middle between the predictions of
$\alpha_B^{(1)}(Q^2)$ and $\alpha_B^{(2)}(Q^2)$, indicating the
self-consistency of our results.
At the origin $Q=0$, our results differ from that obtained by Shirkov 
and Solovtsov \cite{Shirkov97} by a small multiplicative factor 
$\kappa^{-1}_e=0.38$, i.e. $\bar \alpha^{(1)}(0)=0.38\;
\alpha^{(1)}_{\mathrm{an}}(0)$, on the other hand we find that the 
estimates of the background coupling constants $\alpha^{(1)}_B(0)$ 
and $\alpha^{(2)}_B(0)$ are much closer to our predictions than to 
$\alpha^{(1)}_{\mathrm{an}}(0)$. 
Table(\ref{T4}) lists our results for $\bar\alpha^{(1)}(0)$ 
together with $\alpha^{(1)}_{\mathrm{an}}(0)$, 
$\alpha^{(1)}_B(0)$ and $\alpha^{(2)}_B(0)$ for $0\le n_f\le 5$. 
\TABLE[ht]{\centerline{
\begin{tabular}{|c|c|c|c|c|c|c|c|}
\hline
\small
$          n_f      $  &    0  &   1   &   2   &   3   &   4   &   5
\\ \hline
$\bar\alpha^{(1)}(0)$  & 0.432 & 0.460 & 0.492 & 0.529 & 0.571 & 0.620 
\\ \hline
$\alpha^{(1)}
_{\mathrm{an}}(0)  $  & 1.142 & 1.216 & 1.300 & 1.396 & 1.508 & 1.639
\\ \hline
$\alpha^{(1)}_B(0)  $  & 0.448 & 0.448 & 0.448 & 0.446 & 0.427 & 0.391
\\ \hline
$\alpha^{(2)}_B(0)  $  & 0.713 & 0.713 & 0.714 & 0.705 & 0.576 & 0.447
\\ \hline
\end{tabular}}
\caption{The values of $\bar\alpha^{(1)}(0)$ compared to those of
$\alpha^{(1)}_{\mathrm{an}}(0)$, $\alpha^{(1)}_B(0)$ and $\alpha^{(2)}_B(0)$}
\label{T4} }
The numerical value of our coupling constant $\bar\alpha^{(1)}(Q^2)$, as
shown in table(\ref{T4}), is reasonably small which supports the notion of
an expansion in powers of $\bar\alpha^{(1)}$ at low momentum transfers.
Phenomenological verification of this fact would be of large practical
value. Gribov theory of confinement \cite{Close93} demonstrates how colour 
confinement can be achieved in a field theory of light fermions interacting
with comparatively small effective coupling, a fact of potentially great
impact for enlarging the domain of applicability of perturbative ideology to
the physics of hadrons and their interactions \cite{Troyan96}.

In a number of cases of the QCD calculations it is necessary to estimate
integrals of the form \cite{Dokshitzer95,Dokshitzer98Alekseev98}:
\be
F(Q^2)=\int^Q_0\;\alpha_s(k^2)\,\mathrm{f}(k)\,dk\,,
\label{4.6}
\ee
where $\mathrm{f}(k)$ is a smooth function behaving like $k^p$ at $k\ll Q$. 
In this integral, the interval of integration includes the IR region where 
the perturbative expression for the running coupling $\alpha_s(k^2)$ is 
inapplicable. Hence, by introducing an IR matching scale $\mu_I$ 
such that $\Lambda\ll\mu_I\ll Q$ the contribution to integral (\ref{4.6}) 
from the region $k>\mu_I$ can be calculated perturbatively. 
On the other hand, the portion of the integral below $\mu_I$ is expressed 
in Ref.\cite{Dokshitzer95} in terms of a non-perturbative parameter 
$\bar\alpha_p(\mu_I)$ as:
\be
\int^{\mu_I}_0\;\alpha_s(k^2)\,k^p\,dk\,=
\frac{\mu_{I}^{p+1}}{p+1}\:\bar\alpha_p(\mu_I)
\label{4.7}
\ee
For $\mu_I=2$ GeV and $p=0$, an excellent fit to experimental data yields:   
$A(2\,\mathrm{GeV})=\bar\alpha_0(2\,\mathrm{GeV})/\pi=0.18\pm0.03$ 
\cite{Troyan96} and $A(2\,\mathrm{GeV})=0.17\pm 0.01$ \cite{Dokshitzer95}. 
These results agree reasonably well with our estimate for $A(2\,\mathrm{GeV})$,
which is obtained from (\ref{4.7}) by direct substitution of our 
3-flavour formula (\ref{3.28}) for $\alpha_s(k^2)$, giving: 
\be
A(2\,\mathrm{GeV})=\frac{2}{9}\int^2_0\;\frac{1}
{\ln(k^2/\Lambda^2)+E_1(\lambda_e\,k^2/\Lambda^2)}\,dk\,\cong 0.14\,.
\label{4.8}
\ee
Here $\lambda_e=0.04$ and $\Lambda=0.229$ GeV.

\subsection{Gluon condensate}

In QCD instantons are the best studied non-perturbative effects, leading 
to the formation of the gluon condensate, an important physical quantity 
defined as \cite{Shifman79}:
\be
K=\langle 0|\frac{\alpha_s}{\pi}\;G^a_{\mu\nu}G^a_{\mu\nu}|0\rangle\,,
\label{5.1}
\ee
where $G^a_{\mu\nu}$ is the gluon field strength tensor and
$\alpha_s$ is the quark-gluon coupling constant.
This can be related to the total vacuum energy density 
$\epsilon_t$ of QCD by means of the famous trace anomaly 
relation \cite{Gogohia99}:
\be
\theta_{\mu\mu}=-\frac{\beta_0\;\alpha_s}{8\,\pi}\;
G^a_{\mu\nu}G^a_{\mu\nu}+\sum ^{n_f}_q\;m_q\,\overline{\Psi}_q 
\Psi_q+O(\alpha_s^2)\,,
\label{5.2}
\ee
where $m_q$ and $\Psi_q$ denote the quark masses and spinor quark fields  
respectively. Sandwiching $\theta_{\mu\mu}$ between the QCD vacuum states 
in the chiral limit $(\textrm{i.e}.\,\; m_q=0)$ and on account of the 
relation $\langle 0|\theta_{\mu\mu}|0\rangle=4\epsilon_t$ 
\cite{Shifman79,Gogohia99}, one obtains: 
\be
K=-\frac{32}{\beta_0}\;\epsilon_t\,.
\label{5.3}
\ee

In the dilute instanton-gas approximation with $n_f=0$, a direct relation
between the gluon condensate $K$ and the instanton density 
\cite{Shifman79,Bernard79}:
\be
\rho(r)=b\,\Bigg[\frac{2\pi}{\alpha^{(1)}_{\mathrm{PT}}(r^{-2})}\Bigg]^6
\;\exp\Bigg(-\frac{2\pi}{\alpha^{(1)}_{\mathrm{PT}}(r^{-2})}\Bigg)\,,
\label{5.4}
\ee   
with $r\thickapprox 1/Q$ as the instanton scale size variable and $b=0.0015$
\cite{Bernard79}, can be represented in the form \cite{Shifman79}:
\be
K=16\,\int^{\mathrm{R_c}}_0\frac{\rho(r)}{r^5}\;dr\,,
\label{5.5}
\ee   
where $\mathrm{R_c}$ is a cut-off introduced by hand to avoid the
uncontrollable IR divergences.
Since our formula for the running coupling (\ref{3.28})
agrees reasonably well with the estimate of $\alpha^{(1)}_{\mathrm{PT}}$
at short distances $r<\mathrm{R_c}$ and is analytic for all $r\ge 0$,
it is very tempting to use it in place of $\alpha^{(1)}_{\mathrm{PT}}$ 
to evaluate the integral in (\ref{5.5}) in the limit $\mathrm{R_c} \to \infty $.
If we do this, we arrive at:
\be
K=0.25\times11^6\,b\,\Lambda^4\,\int^{\infty}_0\,x^6\,[\,\ln(x^{-2})
+E_1(\lambda_e\,x^{-2})\,]^6\,\exp[-\frac{11}{2}
E_1(\lambda_e\,x^{-2})]\,dx\,,
\label{5.6}
\ee   
here $\Lambda$ corresponds to $n_f=0$. 
A direct numerical integration of (\ref{5.6}) yields:
\be
K=4.8624\;\Lambda^4\,.
\label{5.7}
\ee   
Using the value $\Lambda=0.229$ GeV, which we computed 
in subsection (5.1) for small $n_f$, we estimate the gluon condensate 
from (\ref{5.7}) as $K=0.013\;\mathrm{GeV^4}$. This is in good agreement
with the value phenomenologically estimated from the QCD sum rules approach
\cite{Shifman79}, namely $\langle 0|\frac{\alpha_s}{\pi}\;G^a_{\mu\nu}
G^a_{\mu\nu}|0\rangle\cong 0.012\;\mathrm{GeV^4}$. 
In addition, our estimate for the gluon
condensate is also consistent with that calculated 
phenomenologically in Ref. \cite{Guberina81} and found to be 
$K=0.014^{\,+0.0044}_{\,-0.0018}\;\mathrm{GeV^4}$. 
Note that if we had considered the value 
$\Lambda^{(0)}=0.238$ GeV, obtained from
quenched lattice QCD in Ref. \cite{Capitani99}, instead of our
estimated value of $\Lambda$, we would have then found
$K$ to be 0.015 $\mathrm{GeV^4}$, which is still close 
enough to the two phenomenological estimates mentioned above.
Hence, it may well follow that the instanton density $\rho(r)$ 
is more likely to reveal more realistic and useful information 
when expressed in terms of our analytic running coupling 
$\bar\alpha^{(1)}$.
\FIGURE[ht]{ \unitlength 1cm 
\put(5.9,-0.26){$r\;{\scriptstyle{[\mathrm{GeV}}^{-1}]}$}
\put(13.2,-0.26){$r\;{\scriptstyle{[\mathrm{GeV}}^{-1}]}$}
\put(3.5,-0.6){(a)}\put(11.4,-0.6){(b)}
\put(8.6,5.4){$\bar\rho(r)$}
\put(1,5.4){$\rho_{\mathrm{\scriptstyle{PT}}}(r)$}
\centerline{\resizebox{7cm}{6cm}{\epsfig{file=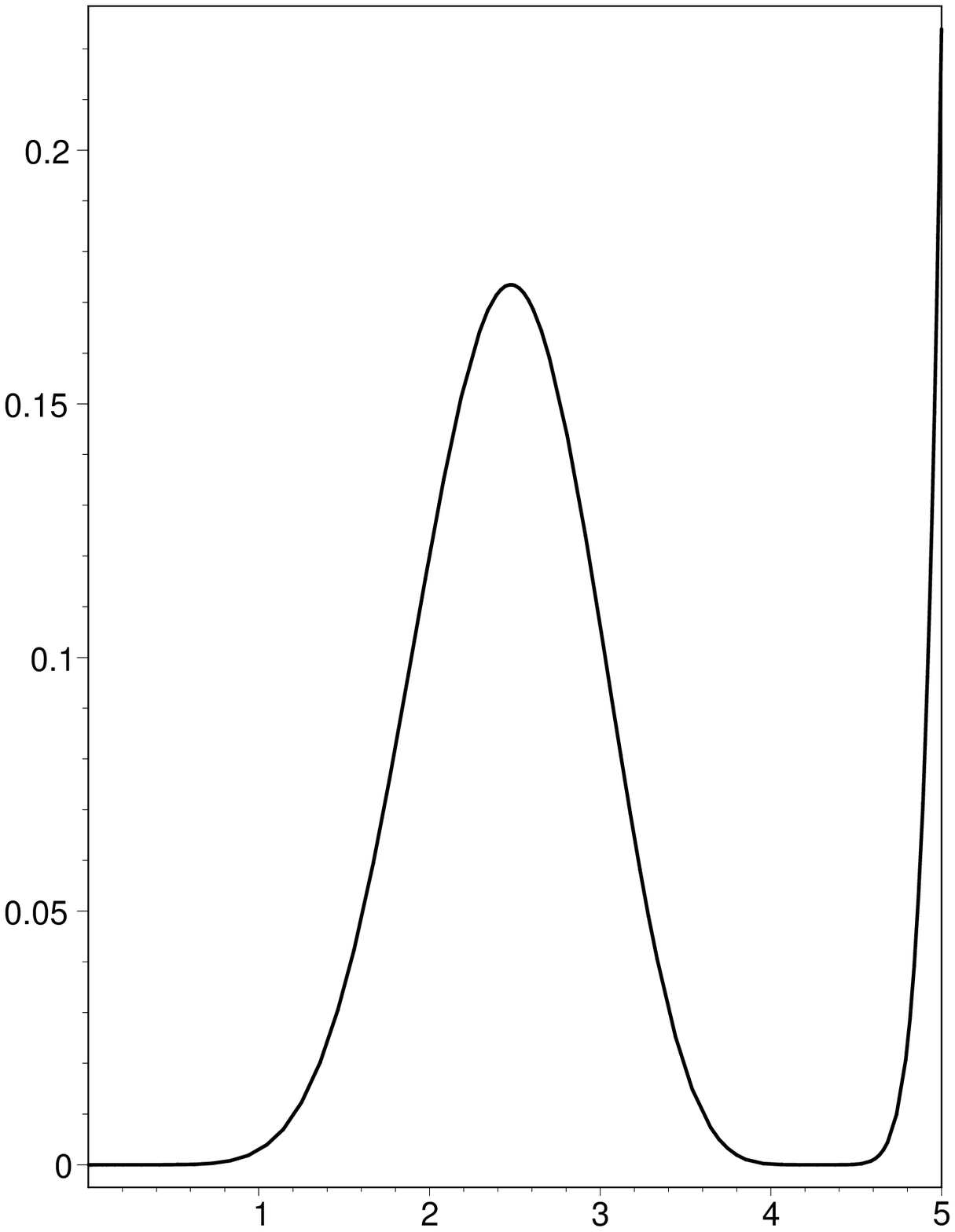}}
\hspace{0.1cm}
\resizebox{7cm}{6cm}{\epsfig{file=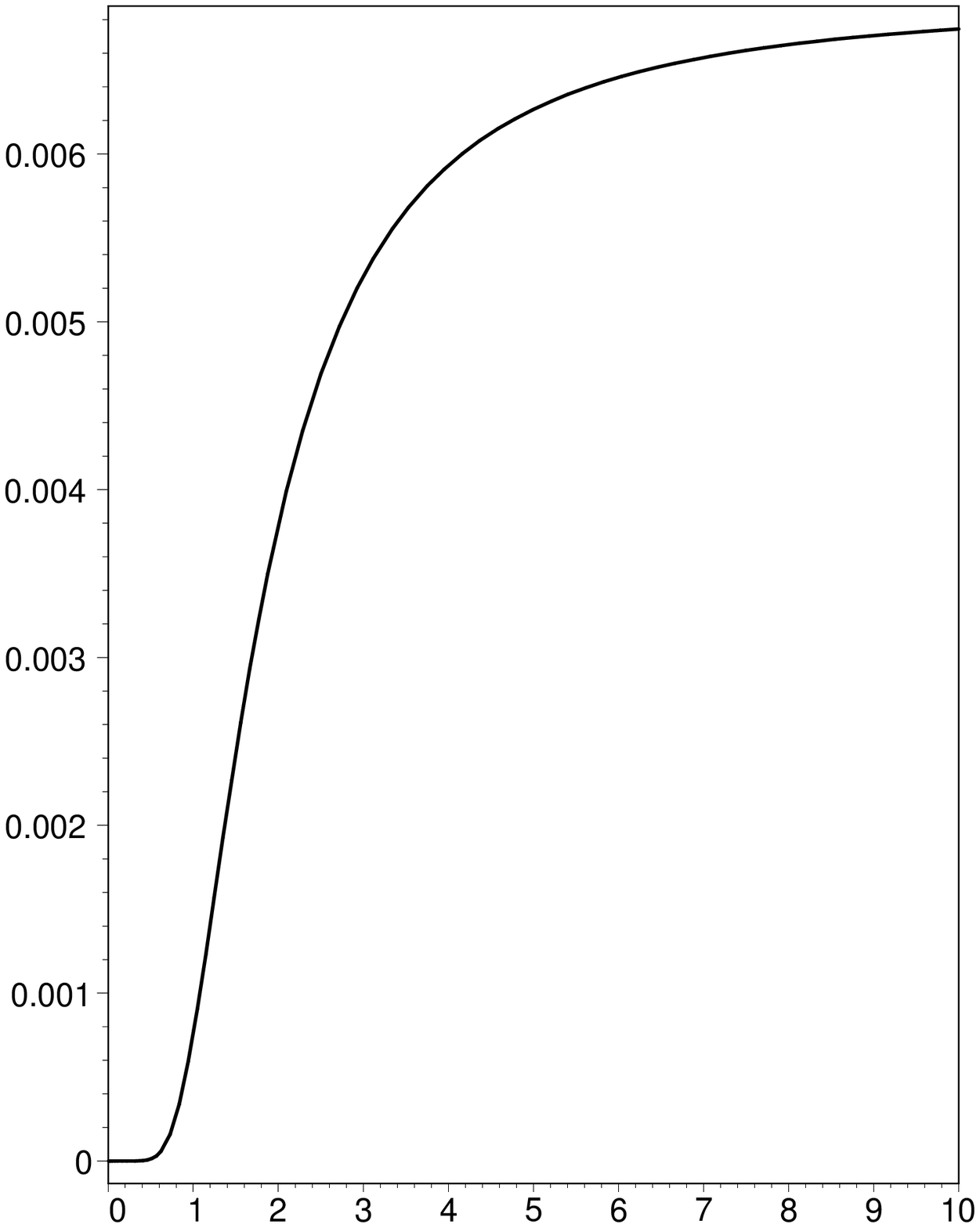}}}
\caption{The behaviour of the instanton density as expressed in terms of 
the running couplings (a) $\alpha^{(1)}_{\mathrm{PT}}$, and (b) 
$\bar \alpha^{(1)}$.}
\label{Fig12} }
In Fig.(\ref{Fig12}), we shed some light on the dependence of
$\rho(r)$ on the running couplings $\alpha^{(1)}_{\mathrm{PT}}$ 
and $\bar\alpha^{(1)}$ which we shall denote by
$\rho_{\mathrm{PT}}(r)$ and $\bar\rho(r)$ respectively. 
As seen in Fig.(\ref{Fig12}), at short distances, say 
$\,r <\, 0.6\;\scriptstyle{\mathrm{GeV}^{-1}}$, $\rho_{\mathrm{PT}}(r)$ 
and $\bar\rho(r)$ agree reasonably well whereas when we increase $r$ to
sufficiently larger values $\rho_{\mathrm{PT}}(r)$ rises very rapidly 
towards infinity while $\bar\rho(r)$ stabilizes to a fixed density.
This reflects the practical usefulness of our approach to the strong
coupling constant.

\subsection{QCD $\beta-$function and IR properties}

In quantum chromodynamics, the renormalisation group $\beta-$function:
\be
\beta(\alpha(Q^2))=Q^2\,\frac{d\,\alpha(Q^2)}{d\,Q^2}\,
,\label{6.1} 
\ee
has the perturbative expansion \cite{Stirling96}:
\be
\beta(\alpha)=-\alpha\,\sum^{\infty}_{n=0}\,\beta_n\,                   
\left(\frac{\alpha}{4\,\pi}\right)^{n+1}\,,
\label{6.2} 
\ee
where $\beta_0$, $\beta_1$ and $\beta_2$ are given by (\ref{1.1}),
(\ref{A3.24}) and (\ref{A3.25}) respectively and \cite{Ritbergen97}:
\be
\beta_3 = 29243.0-6946.30\,n_f+405.089\,n_f^2\,+1.49931\,n_f^3\,,
\label{6.3}  
\ee
here the higher-order coefficients $\beta_n$ for $n>3$ have not been
calculated till now. Generally speaking, (\ref{6.1}) is the crucial equation
that decides whether or not there is a so-called stable fixed point at the 
origin (i.e. whether $\beta(\alpha)=0$ for $\alpha=0$) of either the
infrared or ultraviolet type, with implications for the absence or presence
of asymptotic freedom in the gauge theory under consideration. 
The distinction between an infrared and an ultraviolet stable fixed point
depends on whether the derivative $\beta'(\alpha)>0$ or $\beta'(\alpha)<0$
at $\alpha=0$ respectively. In both kinds of gauge theories (Abelian and 
non-Abelian), there is a stable fixed point at the origin but it turns out
that, while $\beta'(\alpha)>0$ (stable IR fixed point) for an Abelian 
gauge theory, the reverse is true for a non-Abelian gauge theory
, i.e. $\beta'(\alpha)<0$ (stable UV fixed point). In general, it is
extremely difficult to establish the existence of stable fixed points of
a quantum field theory of $\beta(\alpha)$ for $\alpha\ne 0$. 
In perturbative QCD, the one-, two-, three- and four-loop $\beta-$function  
with quark flavours $n_f\le 6$ fail to exhibit any non-trivial zero that
can be interpreted as a stable IR fixed point. Because of this, the 
perturbative QCD coupling constant $\alpha_{\mathrm{PT}}(Q^2)$ blows up as
$Q^2 \to 0$.

Our main task in this section is to demonstrate how the $\beta-$function
in our approach can provide useful information about the IR region that
a truncated perturbative series can not. As a cross check on our method,
we shall calculate the stable IR fixed point for any number of quark
flavour and show that it is consistent with our previous result 
(\ref{3.30}). Starting from (\ref{3.28}) with $\bar\alpha^{(1)}$ being
renamed as $\alpha$, our QCD $\beta-$function in accordance with the
definition (\ref{6.1}) assumes the form:
\be
\bar \beta(\alpha)=-\frac{\beta_0}{4\,\pi}\, \alpha^2 \;
[1-{\mathrm e}^{-W(\alpha)}]\,, 
\label{6.4}     
\ee
where
\be
W(\alpha)={\mathrm e}^{\phi(\alpha)}\,,
\label{6.5}     
\ee
and the function $\phi(\alpha)$ is defined by the transcendental relation:
\be
\phi(\alpha)+E_1(\,{\mathrm e}^{\phi(\alpha)}\,)=\frac{4\,\pi}{\beta_0\,\alpha}
+\ln(\lambda_e)\,.
\label{6.6}     
\ee
In the following domains of $\alpha$:
\be
D_1 = \{ \,\alpha: \phi(\alpha)\ge 3  \, \}
= \{ \, \alpha: 0 < \alpha \le 2.021\,\beta_0^{-1}\, \} \,,
\label{6.7}     
\ee
and
\be
D_2 = \{ \, \alpha: \phi(\alpha)\le -4 \, \} 
= \{ \,\alpha:\, 4.724\,\beta_0^{-1}\le \alpha < 4.757\,\beta_0^{-1}\,\}\,,
\label{6.8}     
\ee
a good approximate solution of (\ref{6.6}) can be found as:
\be
\phi(\alpha) \cong
\begin{cases}
\displaystyle{\frac{4\pi}{\beta_0\alpha}}+\ln(\lambda_e)-
E_1(\lambda_e\,\exp
{ \big( \displaystyle{ \frac{4\pi}{\beta_0\alpha} } \big) } )&
\text{if $\alpha \in D_1$},\\ 
\\
\ln\left(\displaystyle{ \frac{4\pi}{\beta_0\alpha}} - \kappa_e \right)&
\text{if $\alpha \in D_2$}.
\end{cases}
\label{6.9}  
\ee
We note that as $\alpha$ approaches zero, $\phi(\alpha)$ grows without
limit, allowing (\ref{6.4}) to reproduce the conventional one-loop
QCD $\beta-$function. It follows from (\ref{6.4}) that for all $n_f$
satisfying $\beta_0(n_f)>0$ there exists a stable IR fixed point $\alpha_
{\mathrm{FP}}$ identified by $W(\alpha_{\mathrm{FP}})=0$. Using the expansion
of (\ref{2.18}) in (\ref{6.6}), we obtain, in the limit $W(\alpha) \to 0$,
the function:
\be
W(\alpha)=\displaystyle{\frac{4\pi}{\beta_0\alpha}}-\kappa_e \,,
\label{6.10}  
\ee
from which $\alpha_{\mathrm{FP}}$ is found to be:
\be 
\alpha_{\mathrm{FP}}=\displaystyle{\frac{4\pi}{\beta_0\,\kappa_e }}\,.
\label{6.11}  
\ee
This is completely consistent, as it should be, with our previous result
(\ref{3.30}).

For $6<n_f\le 16$, the higher-order terms of the QCD $\beta-$function 
are known to permit the occurrence of IR fixed points. For example, 
for all values of $n_f$ such that $\beta_0(n_f)>0$ and $\beta_1(n_f)
<0$ the two-loop $\beta-$function $\beta^{(2)}_{\mathrm{PT}}(\alpha)$
\footnote{The notations $\beta^{(n)}_{\mathrm{PT}}(\alpha)$ and
$\alpha^{(n)}_{\mathrm{FP}}$ are used to denote the n-loop $\beta-$function
and its corresponding IR fixed point respectively.} 
possesses a non-trivial IR fixed point given by:
\be 
\alpha^{(2)}_{\mathrm{FP}}=-\displaystyle{ \frac{4\pi\,\beta_0}{\beta_1} }
>0\,.
\label{6.12}  
\ee
This may sound fine. However, the IR fixed points arising from the truncation 
of the perturbative series (\ref{6.2}) are likely to be spurious. 
For instance, at the candidate value for $\alpha_{\mathrm{FP}}$, the 
first and second order terms in $\beta^{(2)}_{\mathrm{PT}}(\alpha)$ are 
equal in magnitude (\,i.e. $|\beta_0\,\alpha^{(2)^2}_{\mathrm{FP}}/4\pi|
=|\beta_1\,\alpha^{(2)^3}_{\mathrm{FP}}/(4\pi)^2 |$\,), indicating the 
inefficiency of the two-loop approximation around the calculated 
IR fixed point (\ref{6.12}). In fact, there is no way to prove that 
the IR fixed points extracted from perturbation theory alone are indeed 
the positive zeros of the true $\beta-$function or anywhere near them.
However, it is still worthwhile to test the perturbative expansion
(\ref{6.2}) to 4-loop order against our model for the $\beta-$function.
In Fig.(\ref{Fig13}), we demonstrate the full behaviour of
our $\beta-$function together with its perturbative counterpart 
for $n_f=6$ and 9. As seen in the figure, the $\beta-$function in
our approach $\bar \beta(\alpha)$ agrees very well with the correct 
perturbative predictions at and near the UV fixed point $\alpha=0$. 
As $\alpha$ leaves the region $D$, where $D=\{\alpha: 0\le\alpha\le
2.7/\beta_0(n_f)\,\}$, $\bar \beta(\alpha)$ starts to diverge 
significantly from $\beta^{(1)}_{\mathrm{PT}}(\alpha)$ and changes 
direction, turning back to zero in the limit $\alpha\to\alpha_{\mathrm{FP}}$ 
as depicted in Fig.(\ref{Fig13}.a) for $n_f=6$. The occurrence of IR 
fixed points in $\beta^{(2)}_{\mathrm{PT}}(\alpha)$, $\beta^{(3)}_
{\mathrm{PT}}(\alpha)$ and $\beta^{(4)}_{\mathrm{PT}}(\alpha)$
takes place only if $n_f \in [9,16], [7,16]$ and [8,16] respectively.
In these $n_f-$intervals, we find that the two-, three- and four-loop 
$\beta-$functions behave qualitatively like $\bar \beta(\alpha)$. This 
is illustrated in Fig.(\ref{Fig13}.b) with $\beta^{(3)}_{\mathrm{PT}}
(\alpha)$ and $\beta^{(4)}_{\mathrm{PT}}(\alpha)$ for $n_f=9$. 
It would have been interesting to see whether this qualitatively 
similar behaviour continues beyond the 4-loop order. Unfortunately, 
we know of no existing argument or calculation that definitively 
answers this question. 
\FIGURE[ht]{ \unitlength 1cm 
\put(4,-0.6){(a)}\put(11.4,-0.6){(b)}
\put(4.1,6.2){$\alpha$}\put(11.4,6.2){$\alpha$}
\put(5.5,1.2){$\bar\beta$}
\put(5.4,0.35){$\beta^{(1)}_{\mathrm{PT}}$}
\put(4.2,0.35){$\beta^{(2)}_{\mathrm{PT}}$}
\put(12.25,0.68){$\bar\beta$}
\put(12.3,2.4){$\beta^{(3)}_{\mathrm{PT}}$}
\put(13.8,2.4){$\beta^{(4)}_{\mathrm{PT}}$}
\centerline{\resizebox{7cm}{6cm}{\epsfig{file=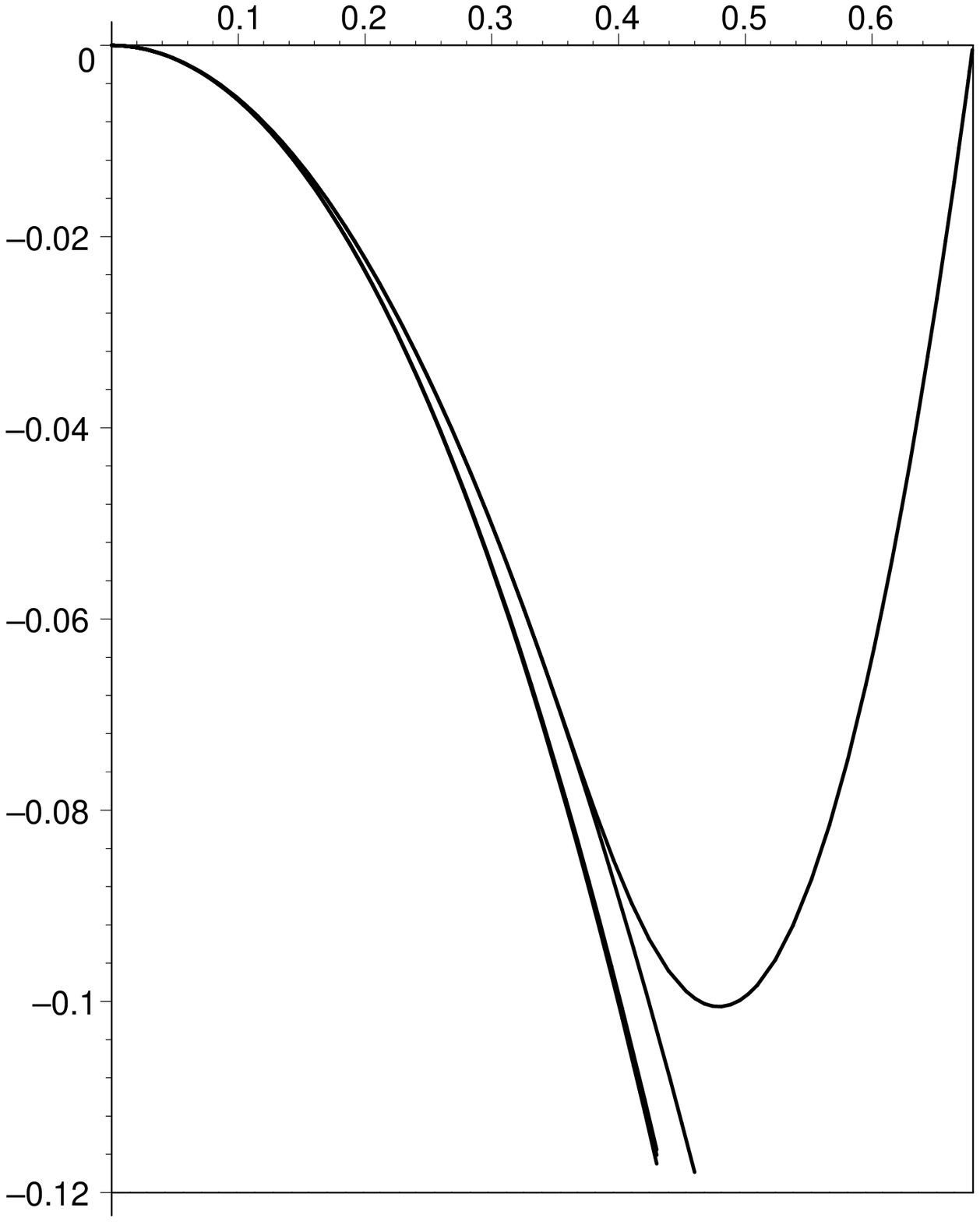}}
\hspace{0.2cm}
\resizebox{7cm}{6cm}{\epsfig{file=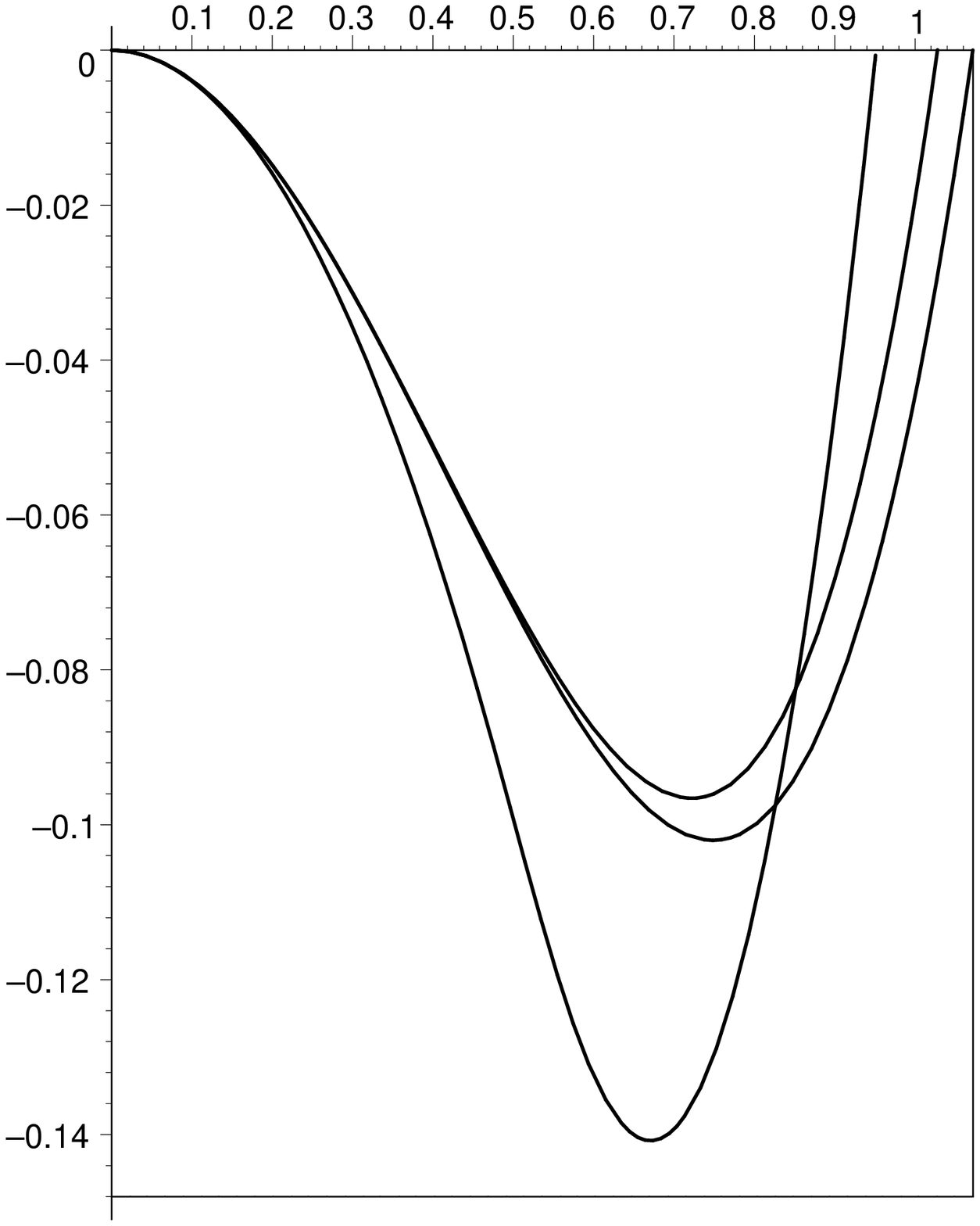}}}
\caption{The behaviour of the $\beta-$function in our approach 
and in perturbation theory for $n_f=6$ (a) and $n_f=9$ (b). In (a), 
the curves of $\beta^{(2)}_{\mathrm{PT}}(\alpha)$,  
$\beta^{(3)}_{\mathrm{PT}}(\alpha)$ and
$\beta^{(4)}_{\mathrm{PT}}(\alpha)$ coincide and none of them
achieves a perturbative IR fixed point. In (b), the locations of 
the fixed points are different but very close to each other. }
\label{Fig13} }
In table(\ref{T5}), we list the values of the IR fixed points in our 
approach and those allowed in perturbation theory within the two-, 
three- and four-loop approximations for $6<n_f<16$. 
\TABLE[ht]{
\begin{tabular}{|c|c|c|c|c|c|c|c|c|c|}
\hline
\small
$         n_f $&   7  &   8  &   9  &  10  &  11  &  12  & 13   &  14  & 15
\\ \hline
$\alpha
_{\mathrm{FP}}$&0.751 &0.839 &0.951 &1.098 &1.297 &1.586 &2.039 &2.854 &4.757 
\\ \hline
$\alpha^{(2)}
_{\mathrm{FP}}$&  *   &   *  &5.236 &2.208 &1.234 &0.754 &0.468 &0.278 &0.143
\\ \hline
$\alpha^{(3)}
_{\mathrm{FP}}$&2.457 &1.464 &1.028 &0.764 &0.579 &0.435 &0.317 &0.215 &0.123
\\ \hline
$\alpha^{(4)}
_{\mathrm{FP}}$ 
               &  *   &1.550 &1.072 &0.815 &0.626 &0.470 &0.337 &0.224 &0.126
\\ \hline 
$\alpha^{(4)}
_{\mathrm{UVFP}}$      

               &  *   &14.364&12.090&5.617 &3.294 &2.295 &1.781 &1.480 &1.286
\\ \hline
\end{tabular}  
\caption{The values of $\alpha_{\mathrm{FP}}$ compared to those of
$\alpha^{(2)}_{\mathrm{FP}}$, $\alpha^{(3)}_{\mathrm{FP}}$ and
$\alpha^{(4)}_{\mathrm{FP}}$. The $*$ sign denotes the non-existence of a
positive fixed point.}
\label{T5}}

For $n_f\in[8,16]$, the four-loop $\beta-$function $\beta^{(4)}_{\mathrm{PT}}$   
has two positive roots. The smaller of these is an IR fixed point whereas
the larger one is an UV fixed point. Thus, we shall denote the latter by
$\alpha^{(4)}_{\mathrm{UVFP}}$. From table(\ref{T5}), we find that with
increasing $n_f$ our IR fixed point $\alpha_{\mathrm{FP}}$ increases
gradually whereas the perturbative roots $\alpha^{(2)}_{\mathrm{FP}}$, 
$\alpha^{(3)}_{\mathrm{FP}}$, $\alpha^{(4)}_{\mathrm{FP}}$ and $\alpha^{(4)}_
{\mathrm{UVFP}}$ decrease to smaller values. However, for most values of $n_f$
we observe that the further $\alpha_{\mathrm{FP}}$ is from one root of some
order, say $\alpha^{(n)}_{\mathrm{FP}}$, the closer it is to another root
of different order, say $\alpha^{(n\pm 1)}_{\mathrm{FP}}$. For example, for 
any $n_f\in[8,10]$ we find $\alpha_{\mathrm{FP}}$ to be closer in magnitude 
to $\alpha^{(3)}_{\mathrm{FP}}$ and $\alpha^{(4)}_{\mathrm{FP}}$ than to
$\alpha^{(2)}_{\mathrm{FP}}$ and $\alpha^{(4)}_{\mathrm{UVFP}}$. On the other
hand, for any $n_f\in[12,14]$ our estimate for $\alpha_{\mathrm{FP}}$ is 
much closer to $\alpha^{(4)}_{\mathrm{UVFP}}$ than to any other value 
of the perturbative IR fixed points. Also, for $n_f=11$ we have $\alpha_
{\mathrm{FP}}\approx\alpha^{(2)}_{\mathrm{FP}}$. This analysis
may not be conclusive however it provides us with some element of truth about
our estimate for the possible magnitude of the true IR fixed point.

It is noteworthy that (\ref{6.11}) enables us to express $\lambda_e$ in terms
of $\alpha_{\mathrm{FP}}$ as:
\be 
\lambda_e=\exp
\left(-\gamma-\displaystyle{\frac{4\pi}{\beta_0\,\alpha_{\mathrm{FP}}}}
\right)\,.
\label{6.13}  
\ee
This could have a direct beneficial effect on our approach if accurate
phenomenological estimates for the IR fixed points were available.
Indeed, if this were the case (\ref{6.13}) would then allow us to 
determine the only free parameter $\lambda_e$ in the theory 
straightforwardly.  
However, in the absence of this phenomenological data, the method we used
in obtaining the value $\lambda_e=0.04$ remains an adequate alternative 
as it results in a good approximation for the QCD $\beta-$function with
any number of quark flavours.


\section{Conclusions}

In this paper, we have developed a new approach for investigating physical 
observables of the type $F(Q^2)$ in regions that are inaccessible to 
perturbative methods of quantum field theory. In this approach, we have
exploited the causality condition to reformulate $F(Q^2)$ as a limit as 
$\lambda\to \infty$ of a contour integral depending on $\lambda$. In this 
way, we have shown that the perturbatively violated analyticity structure 
of physical observables can be reinstated in, at least, the right half
of the complex $Q^2$-plane by taking $\lambda=\lambda_e$ 
instead of $\infty$, where $\lambda_e>0$ is the only free parameter of 
the theory. Explicit guidelines for finding an appropriate value for 
$\lambda_e$ has been given with illustrative examples. We have also shown how
our construction can play the role of a bridge between regions of small and 
large momenta, emphasising the fact that from a {\it{prior}} knowledge 
of either the UV or IR behaviour we can extract information about the 
IR or UV properties respectively. 
In fact, as our formalism incorporates non-perturbative effects
it extends the range of applicability of perturbation theory to cover the 
whole energy domain.

We have demonstrated the implementation of our method in tackling 
the ghost-pole problem in QCD, giving a simple and
new analytic one-loop expression for the strong coupling constant.
As our approach incorporates an extra free parameter $\lambda_e$
in addition to $\Lambda$, we have included a special method
of determining $\lambda_e$ from the UV behaviour of 
the $\lambda$-dependent coupling constant $\bar\alpha^{(1)}(q,\lambda)$.  
This method involves the calculation of the maximum curvature 
of $\bar\alpha^{(1)}(q,\lambda)$ in the UV region $q>1$ 
as explained and carried out numerically in section (4).
In this framework, we have found the effective parameter 
$\lambda_e$ to be 0.04 for $\bar\alpha^{(1)}(q,\lambda_e)$.
Although our prescription of obtaining $\lambda_e$ via the
maximum curvature procedure can be replaced by just tuning
$\lambda_e$ to allow the coupling fit with the experimental 
data available, we find our estimates for $\lambda_e$ appropriate
enough for our model, leading to a good agreement with 
the result $\alpha_{s}(Q^2)=0.38\pm0.03(exp)\pm0.04(theory)$ 
extracted from the fits to charmonium spectrum and fine structure 
splittings \cite{Badalian99} at a low energy scale $Q=1.0\pm0.2$~GeV. 
A supplementary discussion of how to determine the QCD 
scale parameter $\Lambda$ in our model has also been considered.

A distinctive feature of our approach is that
the running coupling freezes to a finite value at the origin
$Q^2=0$, being consistent with a popular phenomenological hypothesis 
\cite{Mattingly94,Stevenson94} namely the IR freezing phenomenon.
This has also been supported by Gribov theory of confinement which 
demonstrates how colour confinement can be achieved in a field theory
of light fermions interacting with comparatively small effective
coupling, a fact of potentially great impact for enlarging 
the domain of validity of perturbative ideology to
the physics of hadrons and their interactions.

For illustration, we have carried out a comprehensive comparison 
between the predictions in our approach and those estimated by 
other theoretical methods. Besides, we have tested our
model on a fit-invariant IR characteristic integral extracted 
from jet physics data \cite{Troyan96,Dokshitzer95} over the energy 
interval $Q\le 2\,\mathrm{GeV}$. This test shows a reasonable 
agreement with our prediction.
For further applications, we have used our model for the running coupling 
to compute the gluon condensate and have shown that our result agrees very
well with the value phenomenologically estimated from QCD sum rules
\cite{Shifman79}. Finally, we have calculated the $\beta-$function
corresponding to our QCD coupling constant and have shown that it behaves
qualitatively like its perturbative counterpart, when calculated beyond
the leading order and with a number of quark flavours allowing for the 
occurrence of IR fixed points.

In further studies, it would be interesting to apply
our analytic approach to improve the standard expressions 
for the running masses in perturbative QCD.


\vskip 1cm

\noindent
{\bf Acknowledgements} I would like to thank Prof. Paul Mansfield for 
useful discussions on this topic.

\vfill
\eject


\end{document}